\documentclass[12pt,titlepage]{article}
\usepackage{graphicx}
\usepackage{natbib}

\usepackage{amsmath}
\usepackage{amssymb}
\usepackage{txfonts}
\usepackage{color}
\usepackage[none]{hyphenat}

\usepackage{authblk}

\title{3D Modelling of the climatic impact of outflow channel formation events on Early Mars.}
\author[1]{Martin Turbet}
\author[1]{Francois Forget}
\author[2]{James W. Head}
\author[3]{Robin Wordsworth}

\affil[1]{Laboratoire de M\'et\'eorologie Dynamique, Sorbonne Universit\'es, UPMC Univ Paris 06, CNRS, 4 place Jussieu, 75005 Paris.}
\affil[2]{Department of Earth, Environmental and Planetary Sciences, Brown University, Providence, RI 02912, USA.}
\affil[3]{Paulson School of Engineering and Applied Sciences, Harvard University, Cambridge, MA 02138, USA.}

\date{\today}

\begin{document}

\maketitle

\begin{abstract}

Mars was characterized by cataclysmic groundwater-sourced surface flooding that formed large outflow channels and that 
may have altered the climate for extensive periods during the Hesperian era. 
In particular, it has been speculated that such events could have induced 
significant rainfall and caused the formation of late-stage valley networks.
We present the results of 3-D Global Climate Model simulations reproducing the 
short and long term climatic impact of a wide range of outflow channel formation events under cold ancient Mars conditions. 
We find that the most intense of these events (volumes of water up to 10$^7$~km$^3$ and released at temperatures up to 320~Kelvins) cannot 
trigger long-term greenhouse global warming, regardless of how favorable are the external conditions (e.g. obliquity and seasons). 
Furthermore, the intensity of the response of the events is significantly affected by the atmospheric pressure, 
a parameter not well constrained for the Hesperian era.
Thin atmospheres (P~$<$~80~mbar) can be heated efficiently because of their low volumetric heat capacity, 
triggering the formation of a convective plume that is very efficient 
in transporting water vapor and ice at the global scale. Thick atmospheres (P~$>$~0.5~bar) 
have difficulty in producing precipitation far from the water flow area, and are more efficient in generating snowmelt.
In any case, outflow channel formation events at any atmospheric pressure are unable 
to produce rainfall or significant snowmelt at latitudes below 40$^{\circ}$N. 
As an example, for an outflow channel event (under a 0.2~bar atmospheric pressure and 45$^\circ$ obliquity) releasing 
10$^6$~km$^3$ of water heated at 300~Kelvins and at a discharge rate of 10$^9$~m$^3$~s$^{-1}$, 
the flow of water reaches the lowest point of the northern lowlands (around~$\sim$ 70$^{\circ}$N, 30$^{\circ}$W) 
after $\sim$~3~days and forms a 200m-deep lake of 4.2$\times$10$^6$~km$^2$ after $\sim$~20~days;
the lake becomes entirely covered by an ice layer after $\sim$~500~days. Over the short term, such an event leaves 6.5$\times$10$^3$~km$^3$ of 
ice deposits by precipitation (0.65$\%$ of the initial outflow volume) and can be responsible for the melting of 
$\sim$~80~km$^3$ (0.008$\%$ of the initial outflow volume; 1$\%$ of the deposited precipitation).
Furthermore, these quantities decrease drastically (faster than linearly) for lower volumes of released water.
Over the long term, we find that the presence of the ice-covered lake has a climatic impact similar to 
a simple body of water ice located in the Northern Plains.

For an obliquity of $\sim$~45$^{\circ}$ and atmospheric pressures $>$~80~mbar, we find that the lake ice is transported 
progressively southward through the mechanisms of sublimation and adiabatic cooling. 
At the same time, and as long as the initial water reservoir is not entirely sublimated 
(a lifetime of 10$^5$~martian years for the outflow channel event described above), 
ice deposits remain in the West Echus Chasma Plateau region where hints of hydrological activity 
contemporaneous with outflow channel formation events have been observed. However, 
because the high albedo of ice drives Mars to even colder temperatures, 
snowmelt produced by seasonal solar forcing is difficult to attain.

\end{abstract}


\section{Introduction}

During the Late Hesperian epoch of the history of Mars (about 3.1-3.6~Gyrs ago; \citet{Hart:01ssr}), 
the large outflow channels observed in the Chryse Planitia area are 
thought to have been carved by huge water floods caused by catastrophic and sudden 
release of groundwater \citep{Bake:82,Carr:96}. 
It has been speculated that such events could have warmed the climate 
and possibly explain the contemporaneous 
formation of dendritic valley networks observed in the nearby Valles Marineris area 
and on the flanks of Alba Patera, Hecates Tholus, and Ceraunius Tholus, 
and that have been interpreted to be precipitation-induced \citep{Guli:89,Guli:90,Bake:91,Guli:97,Guli:98,Guli:01,Mang:04sci,Quan:05,Weit:08,Santiago:12}.
Although the Late Hesperian epoch is thought on the basis of geology and mineralogy 
to have been cold \citep{Head:04,Bibr:06,Ehlm:11},
the characteristics of these dendritic valleys suggest that 
the valleys were formed under persistent warm conditions (.e.g, \citealt{Mang:04sci}).
First, their high degree of branching is interpreted to indicate formation by precipitation.
Second, their high drainage densities - evidence of their high level of maturity - 
and the presence of inner channels favor the presence of 
stable liquid water for geologically long periods of time \citep{Crad:02}. 
Third, sedimentary morphologies observed in the 
region of Valles Marineris \citep{Quan:05} suggest a fluvial and lacustrine environment.
Under this hypothesis, the warm liquid water floods that formed the outflow channels would inject water vapor into the atmosphere, 
a powerful greenhouse gas that could trigger a significant warming period possibly leading to long lasting 
pluvial activity (rainfall).

In this paper, we use a 3-Dimensional Global Climate Model (LMD GCM) to explore the global climatic impact 
of outflow channel water discharge events on a Late Hesperian Mars over a range of temperatures and atmospheric pressures.
These bursts of warm liquid groundwater outflows onto the surface 
can trigger strong evaporation, possibly leading to global climate change.
How warm and how wet was the atmosphere of Late Hesperian Mars after such major outflow channel events?
The climatic effect of relatively small and 
cool groundwater discharges has been studied on a regional scale \citep{Kite:11a} 
and localized precipitation is indicated.
In this contribution, we investigate the climatic impact
at a global scale of a wide range of possible outflow channel events,
including the case of the most intense 
outflow events ever recorded on Mars \citep{Carr:96}.
Our work focuses on both (1) the direct short-term climate change 
due to the initial strong evaporation of water vapor 
and (2) the long-term change of the water cycle due to the 
presence of liquid water and ice at non-stable locations.

When a warm liquid water flow reaches the surface,
strong evaporation occurs and the total evaporation rate increases with
the temperature and the surface area of the flow.
In term of energy budget, a 300K warm liquid water flow can 
potentially convert $\sim$~5$\%$ of its total mass into water vapor before freezing starts.
The injected water vapor will have a major role on the radiative budget of the planet. 
First, water vapor is a greenhouse gas that can absorb ground thermal infrared emission efficiently.
Second, water vapor can condense to form clouds. In the process, large amounts of latent heat can be released in the atmosphere. 
The clouds can reflect the incoming solar flux as well as contribute to an additional greenhouse effect, 
depending on their height and the opacity of the background atmosphere, which depends on the total atmospheric pressure.

To study the global climatic effect of localized outflow channel events,
3D Global Climate Models are particularly relevant because they not only model 
the physical processes described above, but also the 3D dynamical processes that 
play a major role in climatic evolution.
In particular, we show in this paper that 3D dynamical processes (horizontal advection, in particular) 
are key to understanding the relaxation timescale of the Late Hesperian martian atmosphere 
immediately following major outflow channel events.

\section{Background}

\subsection{Outflow channels}

\subsubsection{Description}
\label{outflow_description}

Outflow channels are long (up to~$\sim$~$2000$~km)  and wide (up to~$\sim$~$100$~km) valleys 
that were sculpted by large-scale debris-laden water flows \citep{Bake:82,Bake:92,Carr:96}.
The most prominent martian outflow channels are located in the circum-Chryse area and 
debouch in the Chryse Planitia itself \citep{Carr:96,Ivan:01}.

Several processes have been suggested to have caused such outburst floods \citep{Kres:02jgr}. 
It is likely that the water that was released during these 
events come from subsurface aquifers \citep{Clif:93,Clif:01}.
In this scenario, the temperature of the extracted 
subsurface water is controlled by the geothermal gradient and thus
would depend on its initial depth of origin.
During the Late Hesperian, when outflow 
channel events largely occured, this gradient could have been locally higher \citep{Bake:01},
because the circum-Chryse area is close to the volcanically active Tharsis region \citep{Cass:15}.
Therefore, the discharged water could have reached the surface at a 
maximum temperature of tens of degrees above the freezing point \citep{Kres:02jgr}.
We note that the run-away decomposition of CO$_{2}$ 
clathrate hydrate \citep{Milt:74,Bake:91,Hoff:00,Koma:00}, 
proposed as a possible mechanism for the origin of the outflow water, 
cannot produce water temperature greater than 10K above the freezing point.
To a first approximation, and from a climatic point of view, 
the only difference between these two processes of liquid water discharge
is the temperature of the water. Thus, we considered in this paper various cases ranging from 280 Kelvins to 320 Kelvins (see section~\ref{temp_flow}).

Whatever the physical process operating,
large amounts of water released at very high rates are needed at the origin of the water flow 
in order to explain the erosion of the circum-Chryse outflow channels.
The quantity of water estimated to erode all the 
Chryse basin channels is $\sim 6\times10^6~$km$^3$ 
assuming 40$\%$ by volume of sediment \citep{Carr:96} but could 
possibly be much more if one assume lower sediment loads \citep{Klei:05,Andr:07}, which is, 
for example the case on Earth ($\sim$0.1$\%$ of sediment by volume).

The different estimates of outflow channel single-event volumes, 
discharge rates and durations lead to a wide range of results, but
two endmember scenarios can be defined and explored.
On the one hand, some researchers estimated that only a limited number of very intense 
(volume up to $3\times10^6$~km$^3$, discharge rates up to $10^9$~m$^3$~s$^{-1}$)
outflow channel formation events 
actually occured \citep{Rott:92,Bake:91,Koma:97}.

On the other hand, more recently, other researchers argued that outflow channels were formed 
by numerous individual small events \citep{Andr:07,Harr:08jgr}.
This latter work implies water volumes from hundreds to thousands of km$^3$, 
discharge rates of $10^6$-$10^7$ m$^3$~s$^{-1}$ for individual 
events and a minimum period between successive single events of $\sim$~20 martian years.
These endmember estimates differ by several orders of magnitude, but in this paper, we explored the full range.

\subsubsection{Fate of the outflow channel liquid water flow}

In this section, we provide a description of the possible fate, and 
calculations of the possible velocities, of the outflow channel water; 
these will serve as input for the description of the liquid water runoff 
under various conditions in the GCM simulations.

The ejected liquid water flows from the circum-Chryse 
area all inevitably debouch into the basin of Chryse.
However, Chryse Planitia is not a closed basin and if the total 
amount of water released in a single event is high enough, 
the water will spill into the Northern Plains \citep{Ivan:01}, 
flowing down on slopes inclined at $\sim$0.03$^{\circ}$ for more than 2000km. 
This is an important point because, as the wetted area of the flow increases, 
the total rate of evaporation rises.
The fate of the outflow channel liquid water flow can be subdivided into two steps:

1. First, the ground-surface liquid water flows 'inside' the outflow channels. The Reynolds Number $Re$ of such flows is 
given by  
\begin{equation}
\label{Reynolds}
 Re = \rho U_{c} R_{c} /\mu ,
\end{equation}
with $U_c$ the mean water flow velocity in the channel, $R_c$ the hydraulic radius (see below) of the 
channel, $\rho$ the density and $\mu$ the viscosity of the flow. For most of the outfow channel events, this number must have 
been orders of magnitude higher than 500 \citep{Wils:04}, meaning that the released ground water flows were turbulent.

The most accurate way \citep{Bath:93,Wils:04} to calculate the mean velocity of such flows is to use the Darcy-Weisbach equation: 
\begin{equation}
\label{Uc}
 U_{c}~=~(8 g R_{c} \sin{\alpha} / f_{c})^{1/2}, 
\end{equation}
with $g=g_{\text{mars}}=3.71~m~s^{-2}$ the gravity on Mars, $\alpha$ 
the slope angle of the channel and $f_c$ a dimensionless friction factor which mostly depends on the bed roughness $z_c$ and the 
water depth $h$ of the flow. This factor can be expressed as follows \citep{Wils:04}: 
\begin{equation}
(8/f_c)^{1/2}~=~ a~\log_{10}(R_{c}/z_{c}) + b,
\end{equation}
with $a$ and $b$ two empirical coefficients, which are respectively equal to 5.657 and 6.6303 
if the bed roughness $z_c$ ($z_c$~=~$10^{-2}$~m here) is fixed \citep{Knud:58}:
This leads equation \eqref{Uc} to the following equation:
\begin{equation}
U_{c}~=~(g R_{c} \sin{\alpha})^{1/2}~(a~log_{10}(R_{c}/z_{c}) + b).
\label{Uc_ab}
\end{equation}
The hydraulic radius $R_c$ is defined as the cross-sectional area of the channel divided 
by its wetted perimeter: 
\begin{equation}
R_c \sim (W_c ~ h)/(W_c + 2h), 
\label{Rc}
\end{equation}

with $W_c$ the channel width and $h$ the flow depth.
Because outflow channels are wider than deep ($W_c$~$\sim$~10-100~km wide but $h$~$\leq$~1~km deep), the hydraulic radius $R_c$ can be 
replaced by the depth of the water flow $h$.

To estimate the velocity of the flow according to its discharge rate $Q~=~U_c W_c h$, we solve equation \eqref{Uc_ab} using the 
Lambert special function W defined by $x~=~W(x e^x)$. We obtain:
\begin{equation}
 h~=~ \Bigg(\frac{3~\ln10}{2 a~W_c (g \sin{\alpha})^{\frac{1}{2}}}~\frac{Q}{W\Big( 
\frac{3\times10^{\frac{3b}{2a}}~\ln10}{2a~{z_c}^{\frac{3}{2}}~W_c (g \sin{\alpha})^{\frac{1}{2}}}~Q \Big)}\Bigg)^{\frac{2}{3}}
\end{equation}
 and 
\begin{equation}
\label{vel_eq}
 U_c~=~\Bigg(\frac{2a~(g \sin{\alpha})^{\frac{1}{2}} W( \frac{3\times10^{\frac{3b}{2a}}~\ln10}{2a~{z_c}^{\frac{3}{2}}~W_c (g \sin{\alpha})^{\frac{1}{2}}}~Q )}{3~\ln10~W_c^{\frac{1}{2}}}\Bigg)^{\frac{2}{3}}~Q^{\frac{1}{3}}.
\end{equation}

\begin{figure*}
\centering
\centerline{\includegraphics[scale=0.9]{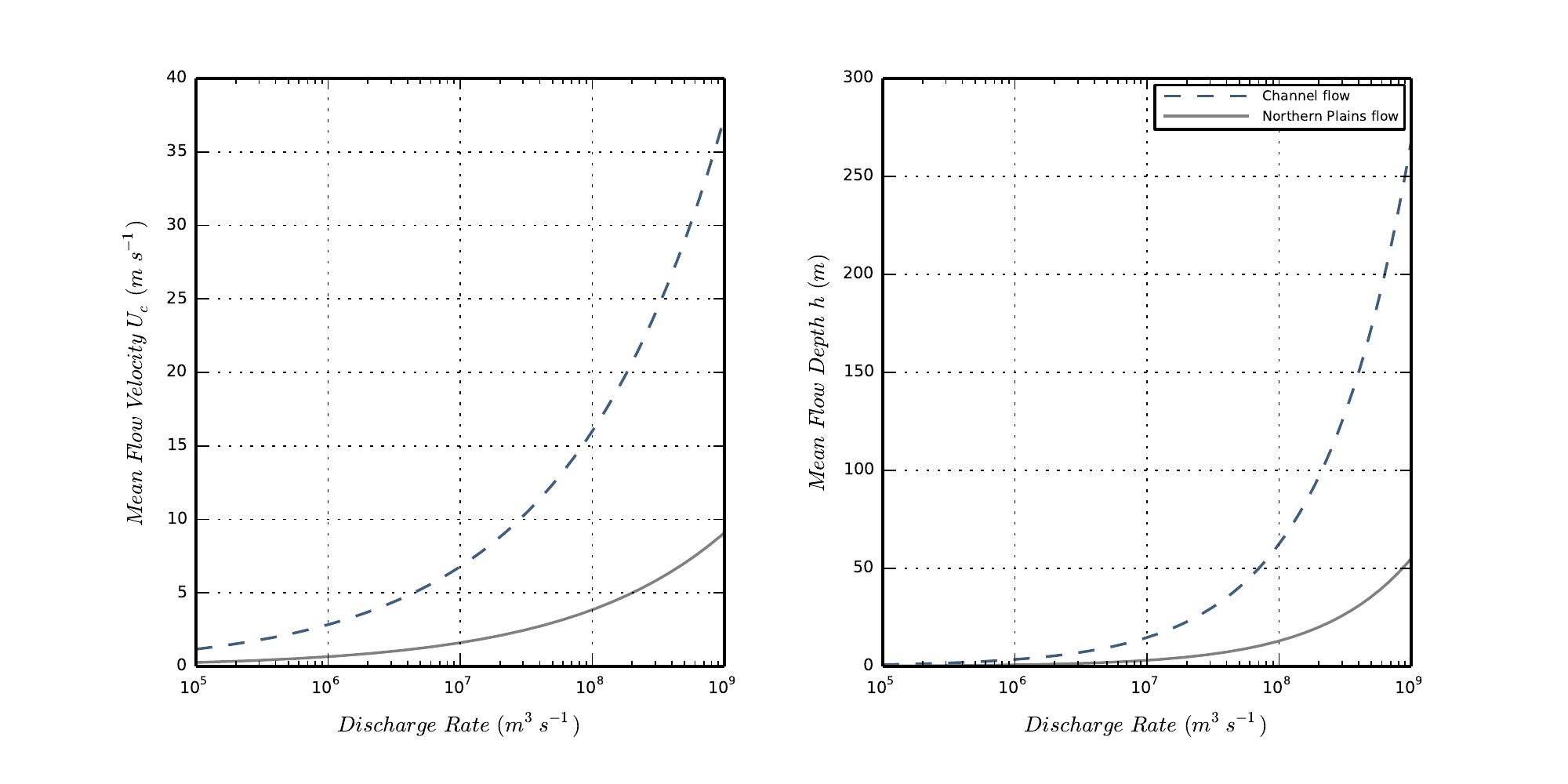}}
\caption{Estimates of mean flow velocity (left) and mean flow depth (right) for 1. (blue) the case of a 100~km-wide 
channel flow on an $\sim$~0.1$^{\circ}$ slope angle and 2. (grey) the case for the same flow spilling onto the Northern Plains 
of Mars ($\sim$~2000~km-wide and slope angle $\sim$~0.03$^{\circ}$). These quantities were calculated for a wide range of water discharge rates, 
using the Darcy-Weisbach equation.}
\label{Flows}
\end{figure*}
The high concentrations of sediments 
in the flows (up to 40$\%$ of the volume) can increase the volumetric 
mass density $\rho$ (initially of $\rho_\text{water}$~$\sim$~1000~kg~m$^{-3}$) by a factor of 2 
and the viscosity $\mu$ (initially of $\mu_\text{water,300K}$~$\sim$~8$\times$10$^{-4}$~Pa~s) by a factor of 16 \citep{Andr:07}, reducing 
by almost 10 the corresponding Reynolds Number. Nonetheless, since both the 
sediment load (from 0.1 to 40~$\%$) and the dependence of the friction factor $f_c$ on the Reynolds Number $Re$, 
are poorly known \citep{Andr:07}, their effects were not taken into account in the flow depth/velocity calculations.

2. As soon as the water flow leaves its channel and reaches Chryse Planitia, the width of the flow strongly increases (up to 2000~km) 
and the slope angle decreases down to 0.03$^{\circ}$. The mean flow velocity and height both decrease (Figure~\ref{Flows}) whereas the wetted area 
increases significantly, leading to even more evaporation. The water will eventually end up in the main topographic depression of Vastitas Borealis 
(around -30$^{\circ}/$70$^{\circ}$ in longitude$/$latitude) building up with time. 
If the volume of water released by the outflow channel event is higher than $\sim$ 2.6$\times$10$^6$~km$^3$, the 
water will spill from the North Polar basin to the Utopia Basin, filling it potentially up to 1.1$\times$10$^6$~km$^3$ \citep{Ivan:01}.
If the volume of water exceeds 3.7$\times$10$^6$~km$^3$, the two basins become connected. 
They can be filled up to few tens of millions of km$^3$.

Once the flow stops, some water will possibly remain in local topographic depressions such as impact craters or 
tectonic basins, thereby contributing to extended evaporation.

If the volume of water or the temperature of the flow are too low, the liquid water flow can potentially freeze before 
reaching the lowest points of the northern lowlands. This would likely occur only for the weakest outflow channel events 
(low volumes/discharge rates/temperatures), and we do not discuss this possibility further in this work.

\subsection{Late Hesperian Climate}
\label{LH_climate}
Late Hesperian Mars was likely to have been cold and dry globally, 
as suggested by the weak occurence of well-developed
valley networks \citep{Carr:96,Fass:08,Harr:05jgr}, 
the absence of observed phyllosilicates within layered deposits 
\citep{Bibr:06,Chev:07b,Cart:13}, 
and the low erosion rates inferred from impact craters morphologies \citep{Crad:93,Golo:06,Quan:15}.

As suggested by the stability of liquid water, and as supported by using the size distribution of ancient craters 
\citep{Kite:14}, the atmosphere of Mars at the end of the Noachian epoch was likely to have been thicker than the 
$\sim$ 8~mbar present day atmosphere.
From the Noachian-Hesperian transition to the Late Hesperian era, 
magmatism may have been responsible for the build up of up to 400~mbar of CO$_{2}$ in the atmosphere \citep[Figure~4]{Grot:11}.
In fact, it is during the period of formation of the outflow channels that 
the release of gaseous CO$_2$ could have been at its maximum \citep{Bake:91,Guli:97}: 
1. Up to 100~mbars of CO$_2$ could have been released by the contemporaneous Tharsis volcanism; 
2. up to 60~mbars of CO$_2$ per volume of 10$^6$~km$^3$ of outflow waters if produced by clathrate destabilization; 
and 3. up to 20~mbars of CO$_2$ per volume of 10$^6$~km$^3$ of outflow waters if coming from 
highly pressurized groundwater reservoirs saturated in CO$_2$.
However, most recent estimates of the several CO$_2$ loss processes (photochemical escape, 
effect of solar wind, sputtering, impact erosion, loss to carbonates, etc.;  
summarized in \citet[section~3]{Forg:13}) suggest that, in spite of the previously mentioned high estimates of 
CO$_2$ outgassing amounts, it is very unlikely that the atmosphere of Late Hesperian Mars 
was thicker than 1~bar. In other words, there are currently no known physical/chemical processes 
that could accommodate the loss of an atmosphere at pressures of more than 1~bar.

To summarize, the Late Hesperian atmosphere was probably thicker than 8~mbar and thinner than 1~bar, 
but the actual surface pressure is still a matter of debate. 
In this paper, we find that the thickness of Late Hesperian Mars atmosphere
plays an important role in relation to the climatic impact of outflow channel formation events. 
We chose to explore a wide possibility of atmospheric surface pressures, ranging from 40~mbar to 1~bar.

\section{Model description}

\subsection{The Late Hesperian Global Climate Model}

In this paper we use the 3-Dimensions LMD Generic Global Climate Model, 
specifically developed for the study of the climate of ancient Mars \citep{Forg:13,Word:13}, 
and adapted here for the study of the influence of outflow channel events on Mars climate during the Late Hesperian.

This model is originally derived from the LMDz 3-dimensional 
Earth Global Climate Model \citep{Hour:06}, 
which solves the basic equations of geophysical fluid dynamics using 
a finite difference dynamical core on an Arakawa C grid.

The same model has been used to study many 
different planetary atmospheres including Archean Earth \citep{Char:13}, a highly irradiated 'future' Earth \citep{Leco:13nat}, 
Pluto \citep{Forg:14agu}, Saturn \citep{Guer:14,Spig:15epsc} and exoplanets \citep{Word:11ajl,Leco:13,Forg:14ptrs,Bolm:16aa,
Turb:16sub1,Turb:16sub2}.

Most of the simulations presented in this paper were performed at 
a spatial resolution of 96~$\times$~48 (e.g. 3.75$^{\circ}$~$\times$~3.75$^{\circ}$; at the equator, 
this gives in average 220~km~$\times$~220~km) in longitude~$/$~latitude.
This corresponds approximately to twice the horizontal resolution used 
and eight times the calculation time needed in the work done by \cite{Forg:13} 
and \cite{Word:13}. For this reason, a parallelized version of the 
GCM was used to deal with the long computation times.
We explored the influence of the horizontal resolution (up to 1$^{\circ}$~x~1$^{\circ}$~$/$~360~$\times$~180 
grid in longitude~$/$~latitude) and did not find any significant discrepency compared with the 96~$\times$~48 lower resolution simulations.

In the vertical direction, the model is composed of 15 distinct atmospheric layers, 
generally covering altitudes from the surface to $\sim$~50~km. 
Hybrid $\sigma$ coordinates (where $\sigma$ is the ratio between 
pressure and surface pressure) and fixed pressure levels were 
used in the lower and the upper atmosphere, respectively. 
The lowest atmospheric mid-layers are located 
around [18, 40, 100, 230, ..]~meters and the highest at about 
[.., 20, 25, 35, 45]~kilometers.

We used the present-day MOLA (Mars Orbiter Laser Altimeter) Mars surface topography \citep{Smit:99,Smit:01mola}, 
and we considered that most of the Tharsis volcanic load was largely in place by the end of the Hesperian epoch \citep{Phil:01}.

We set the obliquity of Mars at 45$^{\circ}$ to be consistent with both the most likely obliquity 
(41.8$^{\circ}$) for ancient Mars calculated by \cite{Lask:04} and 
one of the reference obliquities (45$^{\circ}$) used in \cite{Word:13}.
The sensitivity of obliquity (and more generally of the seasonal effects) is discussed in section~\ref{reference_results}.

To account for the thermal conduction in the subsurface, 
we use an 18-layer thermal diffusion soil model that originally derives from \citealt{Hour:93} 
and has been modified to take into account soil layers with various conductivities. The mid-layer depths 
range from d$_0$~$\sim$~0.1~mm to d$_{17}$~$\sim$~18~m, following the power law 
d$_n$~=~d$_0$~$\times$~2$^n$ with $n$ being the corresponding soil level, chosen to take 
into account both the diurnal and seasonal thermal waves.
We assumed the thermal inertia of the Late Hesperian martian regolith to be 
constant over the entire planet and equal to 250~J~m$^{-2}$~s$^{-1/2}$~K$^{-1}$. 
This is slightly higher than the current Mars global mean thermal inertia in order to account 
for the higher atmospheric pressure.

Subgrid-scale dynamical processes (turbulent mixing
and convection) were parameterized as in \cite{Forg:13} and \cite{Word:13}. 
The planetary boundary layer was accounted for by the \cite{Mell:82} and \cite{Galp:88} 
time-dependent 2.5-level closure scheme, and 
complemented by a ‘‘convective adjustment’’ which rapidly mixes 
the atmosphere in the case of unstable temperature profiles (see section~\ref{convadj} for more details).

In the simulations that include outflow channel events, the dynamical time step is $\sim$~45~seconds 
(respectively $\sim$ 184~s for the control simulations). 
The radiative transfer and the physical parameterizations 
are calculated every $\sim$~15~minutes and  $\sim$~4~minutes 
(respectively every $\sim$~1~hour and $\sim$~15~minutes for the control simulations).

\begin{table}
\centering
\begin{tabular}{ll}
   \hline
              &              \\
   Physical parameters & Values \\
              &              \\
   \hline
              &              \\
   Mean Solar Flux & 465~W~m$^{-2}$ (79$\%$ of present-day) \\
   Obliquity & 45$^{\circ}$ \\
   Bare ground Albedo & 0.2 \\
   Liquid water Albedo  & 0.07 \\
   H$_2$O and CO$_2$ ice Albedos  & 0.5 \\
   Surface Topography & Present-day \\
   Surface Pressure & 0.2~bar \\
   Surface roughness coefficient & 0.01 m \\
   Ground thermal inertia & 250 J~m$^{-2}$~s$^{-1/2}$~K$^{-1}$ \\
              &              \\
   \hline

\end{tabular}
\caption{Physical Parameterization of the GCM.}
\end{table}

\subsubsection{Radiative Transfer in a CO$_2/$H$_2$O mixed atmosphere.}

The GCM includes a generalized radiative 
transfer for a variable gaseous atmospheric composition made of a mix of CO$_{2}$ 
and H$_{2}$O (HITRAN 2012 database, \citet{Roth:13}) 
using the 'correlated-k' method \citep{Fu:92}) 
suited for fast calculation.
For this, we decomposed the atmospheric Temperatures $/$ Pressures $/$ Water Vapor Mixing Ratio 
into the following respective 7~x~8~x~8 grid: 
Temperatures = $\{$100,150,~..~,350,400$\}$~K; 
Pressures = $\{$10$^{-6}$,10$^{-5}$,~..~,1,10$\}$~bar;
H$_2$O Mixing Ratio = $\{$10$^{-7}$,10$^{-6}$,~..~,10$^{-2}$,10$^{-1}$,1 $\}$
~mol of H$_2$O $/$ mol of air (H$_2$O+CO$_2$ here).

CO$_2$ collision-induced absorptions \citep{Grus:98,Bara:04,Word:10co2} ) were included in our calculations as in \cite{Word:13}, 
as well as the H$_2$O continuums. For this, we used the CKD model \citep{Clou:89} with H$_2$O lines truncated at 25~cm$^{-1}$.

For the computation, we used 32 spectral bands in the thermal infrared and 35 in the visible domain.
16 non-regularly spaced grid points were used for the g-space integration, where g is the cumulative 
distribution function of the absorption data for each band.
We used a two-stream scheme \citep{Toon:89} to take into account the radiative effects 
of aerosols (CO$_2$ ice and H$_2$O clouds) and the Rayleigh scattering (mostly by CO$_2$ molecules), 
using the method of \cite{Hans:74}.

In summary, compared to the radiative transfer calculation used in \cite{Word:13}, 
we utilized here a more recent spectroscopic database (HITRAN2012 instead of HITRAN2008) and built 
new correlated-k coefficients suited for wet atmospheres (water vapor VMR up to 100$\%$). In practice, the maximum water 
vapor Mass Mixing Ratio that was reached in our simulations (in the case of low surface pressure simulations) was $\sim$~20$\%$.

In addition, we chose a mean solar flux of 465~W.m$^{-2}$ 
(79$\%$ of the present-day value of Mars; 35$\%$ of Earth's present-day value; 
and 105$\%$ of the flux used in the \cite{Word:13} work), corresponding to the reduced luminosity 
from standard solar evolution models \citep{Goug:81} 3.0 Byrs ago, during 
the Late Hesperian era. During this epoch, the Sun was also 1.5 $\%$ cooler \citep{Bahc:01}; 
we did not, however, include in our model the resulting shift in the solar spectrum.

It is worth nothing anyway that absolute ages are based here on crater counting and are therefore not well constrained.
For instance, the valley networks observed in West Echus Chasma Plateau are 2.9 to 3.4 billion years old \citep{Mang:04sci}.

\subsubsection{CO$_2$ and Water cycles}

Both CO$_2$ and H$_2$O cycles are included in the GCM used in this work. 

1. Carbon Dioxide is here the dominant gaseous species. In our model, 
CO$_2$ can condense to form CO$_2$ ice clouds and surface frost if the temperature 
drops below the saturation temperature. Atmospheric CO$_2$ ice particles are sedimented 
and thus can accumulate at the surface. The CO$_2$ ice layer 
formed at the surface can sublimate and recycle the CO$_2$ in the atmosphere.
The CO$_2$ ice on the surface contributes to the surface albedo calculation: 
if the CO$_2$ ice layer overpasses a threshold value of 1~mm thickness, 
then the local surface albedo is set immediately to the albedo of CO$_2$ ice (0.5 in this work).

2. A self-consistent H$_2$O water cycle is also included in the GCM. 
In the atmosphere, water vapor can condense into 
liquid water droplets or water ice particles, depending 
on the atmospheric temperature and pressure, forming clouds.
At the surface, because the range of surface pressures modeled in this work are well above the 
triple point 6~mbar pressure, liquid water and water ice can coexist. 
Their contributions are both taken into account in the albedo calculation as in \cite{Word:13}.

The stability of liquid water~$/$~ice~$/$~CO$_2$ ice at the surface is governed by the balance 
between radiative and sensible heat fluxes (direct solar insolation, thermal 
radiation from the surface and the atmosphere, turbulent fluxes) and thermal 
conduction in the soil. Melting, freezing, condensation, evaporation, sublimation and 
precipitation physical processes are all included in the model.

\subsubsection{Convective Adjustment}
\label{convadj}

Outflow channel events result in the emplacement of warm liquid water, 
which leads to the sudden and intense warming of the atmosphere.
Global Climate Models ($\sim$~200~km grid size for our simulations) are not suited 
to resolve the convection processes as is done 
in the case of mesoscale models, which have a typical km-size resolution \citep{Kite:11a,Kite:11b}.

Moist convection was taken into account following a moist convective adjustment that originally derives from the 'Manabe scheme'  \citep{Mana:67,Word:13}. 
In our scheme, relative humidity is let free and limited to 100$\%$, since it is 
inappropriate here to use an empirical value for relative humidity (versus altitude) 
that comes from Earth observations, as proposed in the original scheme. 

This scheme has been chosen instead of more refined ones because it is: 1. robust for a wide range of pressures; 
2. energy-conservative; and 3. it is the most physically consistent scheme for exotic 
(non Earth-like) situations such as the ones induced by outflow channel events.
In practice, when an atmospheric grid cell reaches 100$\%$ saturation and the corresponding atmospheric column has 
an unstable temperature vertical profile, the moist convective adjustment scheme is performed to get a stable moist-adiabatic lapse rate.

In our simulations, after major outflow channel events, large amounts of water vapor 
can be released into the atmosphere and the water vapor can easily become a dominant atmospheric species.
In fact we recorded up to 20$\%$ water vapor Mass Mixing Ratios 
following intense outflow channels (in the case of low surface pressure).
Thus, we used a generalized formulation of the moist-adiabiat lapse rate developed by \cite{Leco:13nat} (Supplementary Materials) 
to account for the fact that water vapor can become a main species in our simulations.

In our model we also used the numerical scheme proposed by \cite{Leco:13nat} to account for atmospheric 
mass change after the condensation or the evaporation of gases (water vapor in our case); 
this calculation is usually neglected in most of the well-known Global Climate Models. 
More details on the scheme can be found in \cite{Leco:13nat} (Supplementary Materials).
This scheme comes from previous work for the CO$_2$ cycle on present-day Mars \citep{Forg:98}, where there is some observational validation.

\subsubsection{Parameterization of the precipitation events}

H$_2$O precipitation events were parameterized using a simple cloud water content threshold scheme \citep{Eman:99} as 
in \cite{Word:13}. If the cloud water content overpasses a threshold l$_{0}$ in a given atmospheric grid cell, 
precipitation occurs. We chose l$_{0}$ to be constant and equal to 0.001 kg/kg as in \cite{Word:13}. 
\cite{Word:13} examined the influence of l$_{0}$ and found it to be very low 
(1K difference between l$_0$=0.001 and 0.01 kg/kg). 

We note that the reevaporation of the precipitation is also taken into account in our numerical scheme.

\begin{figure*}
\begin{center}
\centerline{\includegraphics[scale=0.6]{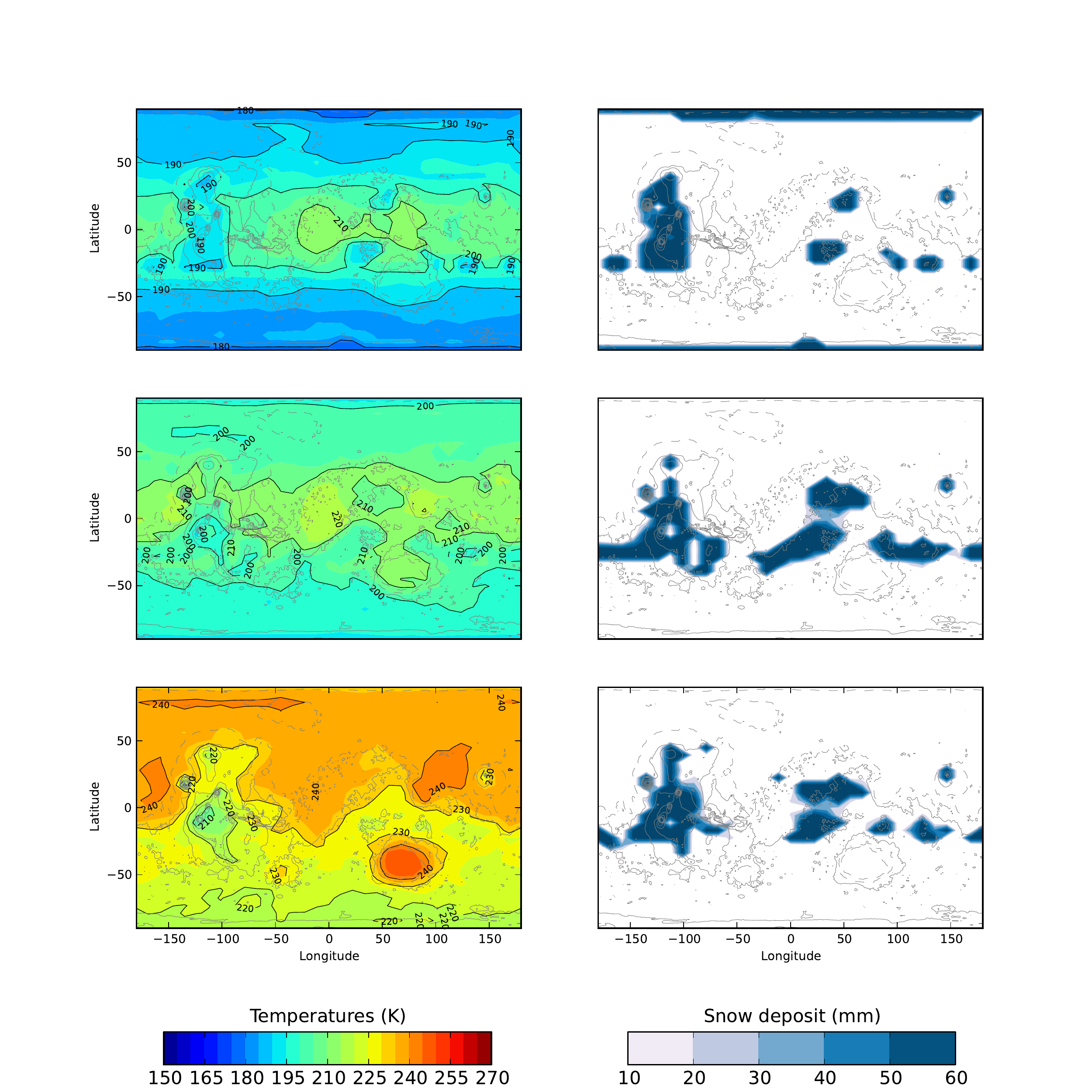}}
\caption{Surface Temperatures (left) and H$_2$O ice deposit (right) 
annual means for the control simulations (at a 96~x~48 horizontal resolution) after $\sim$~800 martian years 
(or 30 loops of the ice iteration scheme). 
Top, middle and bottom plots correspond respectively to control simulations performed at 40~mbar, 0.2~bar and 1~bar.
Grey contours show the topography used in the simulation.
Ice iteration was performed every 2 years, with a 100-year timestep 
used for the first five iterations and 10-year timesteps used thereafter. }
\label{control_simulation}
\end{center}
\end{figure*}

\subsection{Control Simulations without outflow events}
\label{control_simu_section}

We performed control simulations in the conditions described above for 5 different surface pressures 
(40~mbar, 80~mbar, 0.2~bar, 0.5~bar, 1~bar) and we obtained results which are consistent with \cite{Word:13} and \cite{Forg:13}.
For these control runs, the three main differences between our work and \cite{Word:13} were: 1. the updated absorption coefficients 
(now HITRAN 2012); 2. an increase of the solar luminosity (now 79$\%$ of Mars present-day value); and 3. the increase of the horizontal 
model resolution (from 32~x~32 to 96~x~48 in longitude~x~latitude).

Figure~\ref{control_simulation} shows the mean annual surface temperatures and the position of the stable ice deposits for the 
reference case (0.2~bar) and the two surface pressure endmembers (40~mbar and 1~bar).
The mean annual surface temperatures are slightly lower than in Figure~3 in \cite{Word:13} 
which were obtained for a fixed 100$\%$ relative humidity. It is also perhaps due to a slightly reduced CO$_2$ ice cloud 
warming effect at high spatial resolution. 
The stable surface ice deposit locations were calculated using the ice equilibration algorithm of \cite{Word:13}. 
Starting from a random initial surface ice distribution, (1) we run the GCM for two martian years then 
(2) we extrapolate the ice layer field $h_{\text{ice}}$ evolution calculation using:
\begin{equation}
h_{\text{ice}}(t+n_{\text{years}})=h_{\text{ice}}(t)+n_{\text{years}}~\times~\Delta h_{\text{ice}},
\label{ice_iteration_scheme}
\end{equation}
with $\Delta h_{\text{ice}}$ the annual mean ice field change of the one-martian-year previous simulation and 
$n_{\text{years}}$ the number of years requested for the extrapolation.
Then, (3) we eliminate of seasonal ice deposit and (4) we normalize the extrapolated ice field by the initial ice 
inventory to conserve the total ice mass.
Eventually, (5) we repeat the process.

This algorithm has been shown \citep{Word:13} to be insensitive to the proposed
initial ice field location at the beginning of the simulation, at least assuming that the scheme 
has been repeated a sufficient number of times.

In total, for our control simulations, we performed the scheme 30 times, with $n_{\text{years}}$=100 for the first 5 loops and 
$n_{\text{years}}$=10 for 20 more loops for a resolution of 32~x~32. Then, we ran the algorithm 5 more times at the increased resolution 
of 96~x~48 to obtain a stable initial state necessary for the implementation of outflow channel events.

We note that 3D climate modeling under conditions similar to those described above \citep{Forg:13,Word:13} 
have not yet been able to produce liquid water or at least significant precipitation by climatic processes
anywhere on the planet, even when maximizing the greenhouse effect of CO$_{2}$ ice clouds.

\subsection{Experiment - Modeling of Outflow Channel Events}

\subsubsection{Description of the parameterization}
\label{param_outflow}

\begin{figure*}
\begin{center}
\centerline{\includegraphics[scale=0.4]{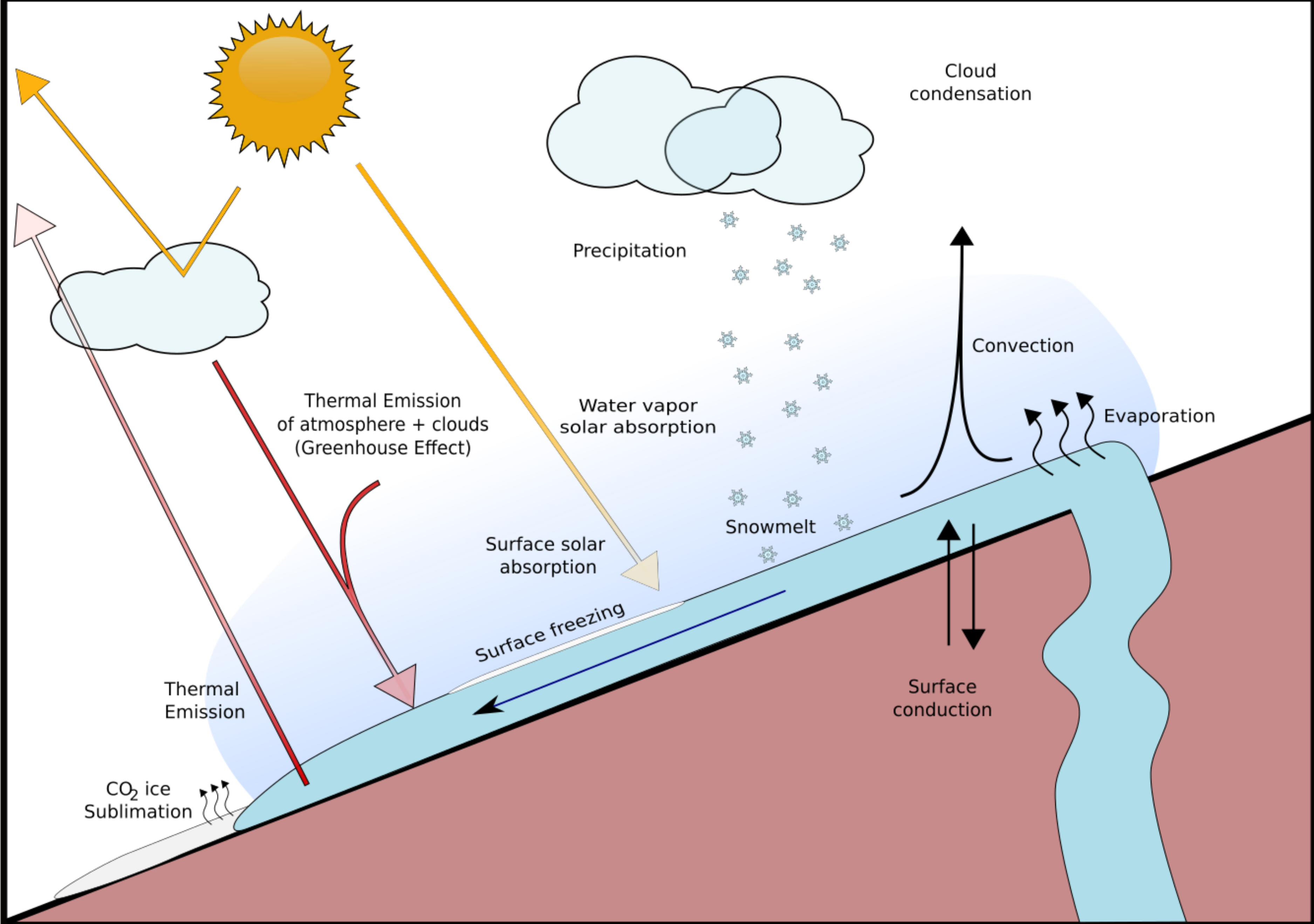}}
\caption{Schematic drawing of the physical processes taken into account during outflow channel events in our GCM simulations.}
\label{Model}
\end{center}
\end{figure*}

Outflow channel events can be modeled to a first approximation by the sudden release, 
and then the spread of warm liquid water over the surface of Mars.
In our simulations, this was accomplished by the emplacement of a fully mixed layer of warm liquid water at the surface. 
The fate of this water depends on the following processes (summarized in Figure~\ref{Model}).:

1. The liquid water layer loses some energy by thermal conduction to the initially cold ground. 
For this, we fix the uppermost of the 18th martian regolith layers at the temperature of the water, and calculate the 
heat flux lost (or gained) by the warm water to the downward layers.

2. The warm liquid water layer cools by emitting thermal infrared radiation at $\sigma$~T$_{\text{surf}}^4$. 
This emission contributes to the radiative transfer budget.

3. The liquid water evaporates and looses some latent heat. The evaporation $E$ at the location of the warm water was computed 
within the boundary-layer scheme, using the following bulk aerodynamic formula:
\begin{equation}
E~=~\rho_{1} C_d V_{1} [q_{\text{sat}}(T_{\text{surf}})-q_{1}],
\label{evap}
\end{equation}
where $\rho_{1}$ and V$_{1}$ are the volumetric mass of air and the wind velocity at the first atmospheric level, 
$q_{\text{sat}}(T_{\text{surf}})$ is the water vapor mass mixing ratio at saturation at the surface, 
and $q_{1}$ is the mixing ratio in the first atmospheric layer. The aerodynamic coefficient is given by 
$C_d$~=~${(\kappa / ln(1+z_1 / z_0))}^2$~$\sim$~2.5$\times$10$^{-3}$, where $\kappa$~=~0.4 is the Von Karman constant, $z_0$ is the roughness coefficient and 
$z_1$ is the altitude of the first level ($\sim$~18~meters).

We modeled the sensible heat exchanged between the surface and the first atmospheric layer using a similar formula: 
\begin{equation}
F_{\text{sensible}}~=~\rho_{1} C_p C_d V_{1} [T_{\text{surf}}-T_{1}], 
\label{sensible}
\end{equation}
with T$_1$ the temperature of the first atmospheric level and C$_p$ the mass heat capacity assumed equal to 850 J~K$^{-1}$~kg$^{-1}$ 
in case of a CO$_2$-only atmospheric composition.

4. Depending on the volume of water modeled, liquid water will flow 
from the Circum-Chryse outflow channel sources to Chryse Planitia, then 
to Acidalia Planitia, and eventually to the Northern Plains.
First, we modeled the displacement of the flow calculated from its height and its velocity. The velocity of the flow mostly 
depends on its width but also on the slope of the terrain. For each grid, we used the subgrid mean slope and 
the subgrid mean orientation of the slope to evaluate (using equations \eqref{Uc_ab} and \eqref{vel_eq})
the velocity and the direction of the flow. Second, we used 
a simple bucket scheme to model the progressive filling of the topographic depressions.

Warm waters flowing on the Northern Plain slopes can also encounter H$_2$O ice (it can be either stable 
at a particular latitude, or related to previous outflow channel events, but from the point of view of 
latent heat exchange and climate, it does not change anything) 
or seasonal CO$_2$ ice (typically present for atmospheres thinner than 1~bar). 
We modeled the interaction of 
H$_2$O and CO$_2$ ices with warm liquid water using energy conservation. If the liquid water is warm and in a sufficient amount, 
all the CO$_2$ ice sublimates and is added to the atmosphere. Similarly, all the water ice 
encountered by the warm flow is melted and converted at the resulting equilibrium temperature.

Once the flow has reached a stable position (e.g. forming a lake), in reality some water may be trapped in local 
topographic depressions (impact craters, tectonic basins, ...); it is difficult, however, to estimate adequately how much 
water might be sequestered in this manner. 
First, the detailed topography of the terrains is unknown prior to resurfacing by the outflow channel events. 
Second, the water outflows themselves modified (and probably smoothed) the topography.
Thus, to take into account not only the effect of the trapped water but also the role of the wet ground, 
we arbitrarily placed a minimum 20~cm layer of liquid water in all the locations where the liquid water flow passed through. 
This assumption may also be representative of the fact that in reality the discharge rate 
does not have a rectangular shape (in time) as we assumed in our parameterizations.

5. As time goes on, the liquid water flow cools. If its temperature 
reaches the 273.15~K freezing temperature (assuming no salts), the water starts to freeze. 
On Earth, salinity drives the freezing point of 
oceans to $\sim$~271K and assuming similar salt rates in outflow waters would not change much our results. 
To account for this process, we developed a multiple layer modified version of the soil thermal conduction model 
already included in the GCM. We have in total 100+ layers, with mid-layer depths 
ranging from d$_0$~$\sim$~0.1~mm to d$_{14}$~$\sim$~2~m, following the power law 
d$_{n,n\leq14}$~=~d$_0$~$\times$~2$^n$ with $n$ being the corresponding soil level and 
the linear law d$_{n,n>14}$=d$_{14}\times(n-13)$ for the deepest layers.
The layers are separated into two parts: the ice cover above and the liquid water below.
For the water ice layers, we use a thermal conductivity of $2.5~$W$~$m$^{-1}~$K$^{-1}$ 
and a volumetric heat capacity of 2$\times$10$^6$~J~m$^{-3}$~K$^{-1}$. 
For the liquid water, we use, respectively, 
a thermal inertia of 20000~J~m$^{-2}$~K$^{-1}$~s$^{-1/2}$ (artificially high to account for convection) 
and a volumetric heat capacity of 4$\times$10$^6$~J~m$^{-3}$~K$^{-1}$.
At each physical timestep, we estimate the thermal diffusion flux lost by the liquid water layer to the water ice layer and 
calculate (using the conservation of energy) the amount of liquid water to freeze. If the depth of the ice - initially going 
down to d=d$_{n}$ - overpasses the layer d$_{n+1}$, we convert the $n+1$ layer into ice.

We note that the use of a multi-layer soil model is important to describe the sea-ice 
formation, evolution and its impact on possible cold early martian climates. 
Such refined models are better suited to represent the temperature 
profile evolution within the ice layer (that may evolve with seasonal forcing or as the ice layer thickens) 
and thus the surface temperature that controls the sublimation rate. In particular, our simulations show 
that up to 95$\%$ of the annual sublimation rate can be produced during the summer seasons. 
This requires a good estimate of the seasonal variations of the surface temperature above the ice.

Simultaneously, as the ice layer forms, we also linearly increase the surface albedo from 
A$_{\text{liq}}$~=~0.07 (if no ice) to A$_{\text{ice}}$~=~0.5 (if the ice layer thickness $h$ overpasses the threshold 
value of $h_*$~=~3.5~cm; \citealt{Letr:91}) as follows: 

\begin{equation}
A = A_{\text{liq}} + (A_{\text{ice}} - A_{\text{liq}})~\frac{h}{h_*}.
\label{lake_ice_5}
\end{equation}

6. The amount of water delivered by outflow events can be very large and thus lead to the accumulation of large quantities of liquid water. 
The timing expected for this water to freeze can be evaluated using a combination 
of the thermal conduction flux in the ice layer 
$F~=~\lambda_{\text{ice}}~\frac{ (T_{\text{surf}}-T_{\text{bottom}})}{h}$ and the conservation of energy. 
Assuming that the temperature in the frozen layer varies linearly between $T_{\text{bottom}}$~=~273.15~K and $T_{\text{surf}}$ (assumed constant) 
as hypothesized in classical 2-layers thermodynamical models \citep{Codr:12}, we have:

\begin{equation}
{\rho}_{\text{ice}}~(L_m~-C_{\text{ice}}~\frac{(T_{\text{bottom}}-T_{\text{surf}})}{2})~\frac{\partial h}{\partial t}~=~\lambda_{\text{ice}}~\frac{ (T_{\text{bottom}}-T_{\text{surf}})}{h}, 
\label{lake_ice_timing}
\end{equation}
where $\rho_{ice}$ is the volumetric mass of the ice ($9.2\times10^2~$kg$~$m$^{-3}$), $C_{\text{ice}}$ is the specific heat capacity of the ice 
($2.1\times10^3~$J$~$kg$^{-1}~$K$^{-1}$), $\lambda_{\text{ice}}$ is the conductivity of the ice ($2.5~$W$~$m$^{-1}~$K$^{-1}$) and 
$L_{m}$~$\sim$~3.34$\times10^5$~J~kg$^{-1}$ is the latent heat of fusion of water ice. 

This leads after integration over time to an expression of t(h), the timing required to freeze a layer of depth h:

\begin{equation}
t(h)~=~\frac{{\rho}_{\text{ice}}}{2~{\lambda}_{\text{ice}}}~(\frac{L_m}{(T_{\text{bottom}}-T_{\text{surf}})}-\frac{C_{\text{ice}}}{2})~h^2.
\label{lake_ice_timing_int}
\end{equation}
For example, the outflow event presented in section~\ref{reference_results} leads to the accumulation of up to 600~meters of liquid water.
A typical timescale (for $T_{\text{surf}}$~$\sim$~200~K) for this water to freeze, according to equation \ref{lake_ice_timing_int}, 
is $\sim$~4~$\times$~10$^3$ martian years.

To account for such long timescales, we developed a modified version of the ice iteration scheme presented above.
(1) First, we run the GCM for a few years then (2) every 2 years, we extrapolate the amount of ice that has locally 
condensed and sublimed in the simulations by 
an arbitrary factor n$_{\text{years}}$. Simultaneously, (3) we proceed to a linear extrapolation of the amount of frozen water/of the 
growth of the ice layer thickness by the same factor n$_{\text{years}}$, using the conservation of energy.
We actually fit the t~=~f(h) function by straight lines of sizes multiple of n$_{\text{years}}$. 
In the reference simulation presented in section~\ref{reference_results}, we 
performed first 5 martian years, then we extrapolated every 2 years using n$_{\text{years}}$=[5,5,20,20,50,50,100,100,500,500].
After the extrapolation of the ice field/the ice layer depth is completed, 
(4) we arbitrarily set the ground temperature profile (where liquid water remains) 
to be linear, between $T_{\text{bottom}}$~=~273.15~K and $T_{\text{surf}}$ calculated using 
the conservation of energy. This is a way to take into account (at first order) the evolution 
of the deepest ground layers that require very long timescales 
to stabilize their temperature profiles. The year following the extrapolation is thus 
also useful to get back a consistent temperature 
profile in the first layers (up to 15~meters typically).

7. Once the outflow water is completely frozen, we use again the ice iteration scheme (see section~\ref{control_simu_section}) 
to get estimates of the timing required for the ice to reach its stable positions. 

\section{Results - the reference simulation}
\label{reference_results}

We present in this section the results of simulations of outflow channel formation events occuring in the largest 
of the Circum Chryse channels: Kasei Vallis.
We chose this particular location because 1. The Kasei Vallis outflow channel begins in Echus Chasma, 
which is close to the West Echus Chasma Plateau valley networks; and 2. Kasei Vallis is one of the largest 
outflow channels on Mars \citep{Carr:96}.

\begin{figure*}
\centering
\centerline{\includegraphics[scale=0.7]{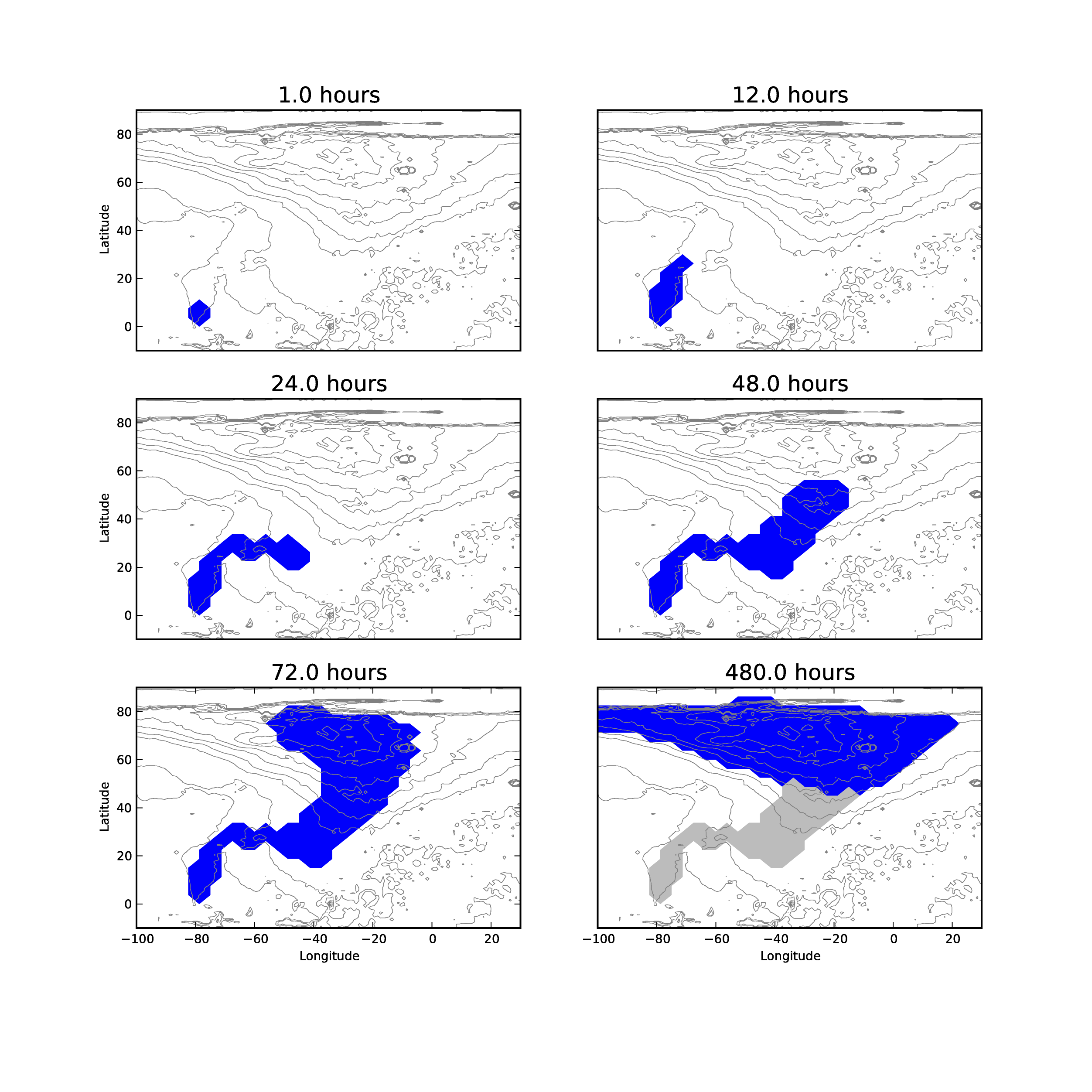}}
\caption{Time lapse of the runoff of the outflow channel event occuring in Echus Chasma, and flowing from Kasei Vallis down to 
the Northern Plains main topographic depression. The blue area corresponds to the position of the flow. The grey color was used to 
represent the 'wet' regions where the flow passed through but did not accumulate.}
\label{runoff}
\end{figure*}

We focus first on a discharge of 10$^6$~km$^3$ (6.9 meters of GEL - Global Equivalent Layer) 
of liquid water heated at 300 Kelvins. 
Water is released at a constant rate of 10$^9$~m$^3$~s$^{-1}$ in the region of Echus Chasma 
(see Figure~\ref{runoff} for the associated flow).
This event is an upper estimate (in volume, discharge rate and temperatures) of 
the characteristics of outflow channel formation events (see section~\ref{outflow_description} for references). 

As explained in section~\ref{LH_climate}, surface atmospheric pressure in the Late Hesperian epoch is poorly constrained. 
Thus, we focus first on the case of a surface pressure of 0.2~bar.

\subsection{Description of the flow}

A volume of 10$^6$~km$^3$ of liquid water is released at the discharge rate of 1~km$^3$~s$^{-1}$. 
It takes approximately 1.1 martian days for the liquid water to travel from the source of the flow 
(in Echus Chasma, at $\sim$ 4$^{\circ}$N,-79$^{\circ}$E) to the end of Kasei Vallis 
(at $\sim$ 30$^{\circ}$N,-45$^{\circ}$E), and 1.5 more days for the same flow to reach the 
main topographic depression of the northern plains (at $\sim$ 70$^{\circ}$N,-30$^{\circ}$E).
This corresponds, respectively, to mean flow speeds of $\sim$~30~m~s$^{-1}$ and 
$\sim$~16~m~s$^{-1}$, which are consistent with the two endmembers values shown in Figure~\ref{Flows}.

After $\sim$~11 days, the source of ground water (located in Echus Chasma) becomes inactive. Eventually, it takes approximately 20 martian days 
in this scenario for the liquid water that has erupted in Echus Chasma to form a stable lake in the 
lowest part of the Northern Plains. This lake extends over an area of 4.2 millions of km$^2$ 
($\sim$~2.9$\%$ of the global surface area of Mars), has a mean depth of $\sim$~240~meters and a peak depth of $\sim$~600~meters.
Some water ($\sim$ 20~centimeters) is left at locations with latitude $<$ 50$^{\circ}$N to 
account for the wet ground and the water possibly trapped in the topographic depressions.

The fate of the outflow channel formation event can be divided into two main parts:

1. During the first $\sim$~500 days following the 
event, the 'Warm Phase', an intense hydrological cycle takes place. 
The end of this phase approximately coincides with the time when the Northern Plains lake becomes fully covered by an ice layer. 

2. During the following $\sim$~10$^5$ martian years, the martian climate is controlled by a weak and cold water cycle. 
It takes approximately the first 4~$\times$~10$^3$ years (as predicted by simple energy-balanced models; \citealt{Kres:02jgr}) 
for the lake to be entirely frozen, 
and the rest to sublimate the lake completely and move the ice to its positions of equilibrium, 
assuming no ice gets buried below a lag deposit or gets transported through glacier flows.

\subsection{The Warm Phase}

As soon as the simulation starts, the warm 300~K liquid water 
released in Echus Chasma evaporates efficiently following equation~\ref{evap}, while flowing 
over the Northern Plains slopes.
\begin{figure*}
\centering
\centerline{\includegraphics[scale=0.85]{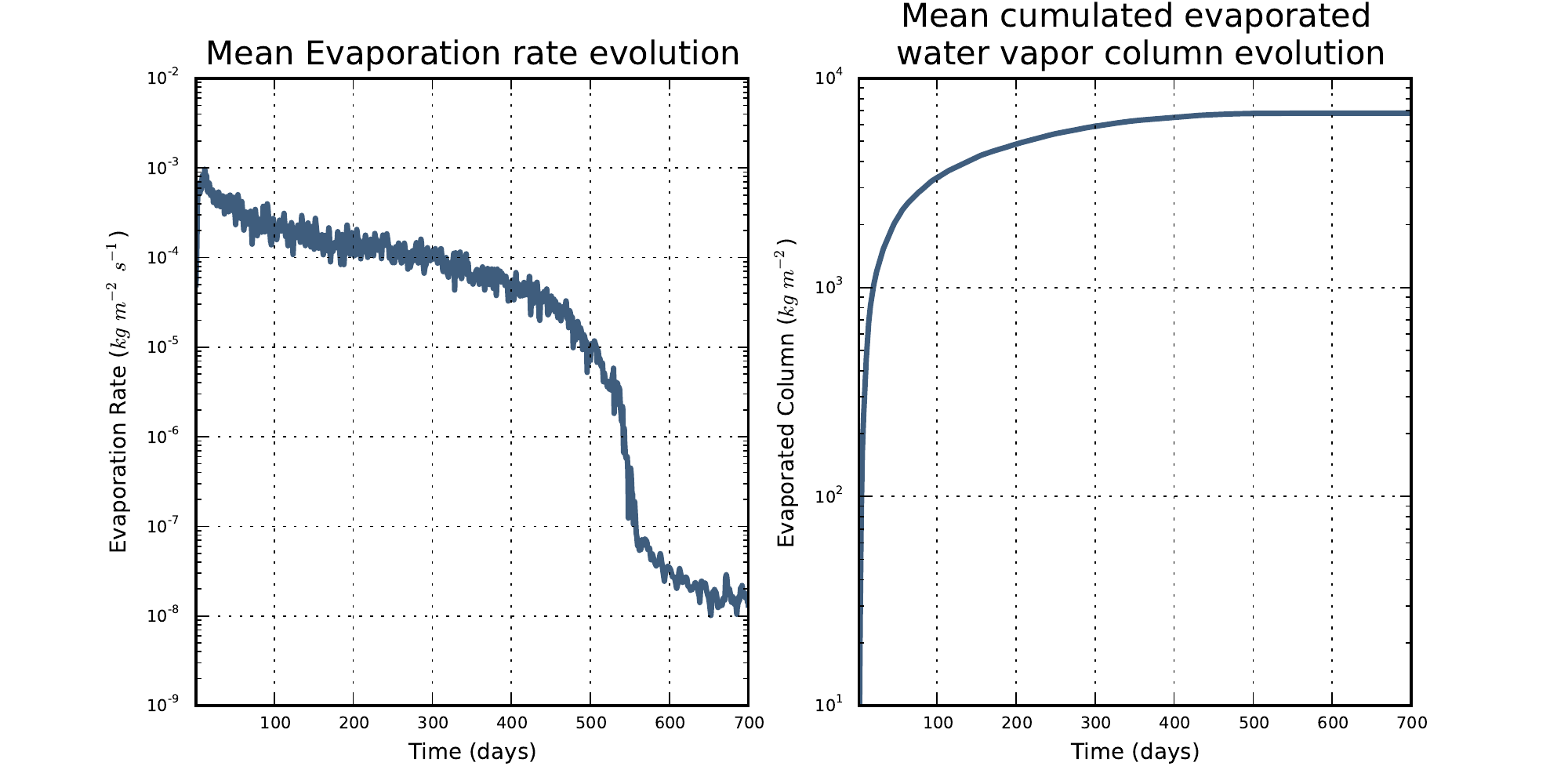}}
\caption{Mean lake evaporation rate evolution (left) and mean cumulative evaporated liquid water from the lake (right). 
The right curve is the cumulative integral over time of the left curve.}
\label{evaporation_fig}
\end{figure*}
At the locations reached by the flow, which represent $\sim$ 11 million km$^2$ ($\sim$~7.5$\%$ of the global surface area of Mars), 
the evaporation rate can reach $\sim$~10$^{-3}$ kg~m$^2$~s$^{-1}$ for tens of days. Figure~\ref{evaporation_fig} (left) shows 
the mean evaporation rate for the 4.2$\times$10$^6$~km$^2$ Northern Plains stable lake formed by the outflow channel flood accumulation. 

During the 500 days following the event, 
a global precipitable water amount of $\sim$~23 centimers is evaporated by the liquid water flow. 
Evaporation of the lake accounts for 96~$\%$ of this amount (blue region in Figure~\ref{runoff}, after 480 hours) 
and 4~$\%$ by the evaporation of the transient flow (grey region in Figure~\ref{runoff}, after 480 hours). 
This amount of cumulative evaporation corresponds to $\sim$~3.4 $\%$ of the initial volume of water ejected 
by the ouflow event, which is approximately 0.7~times the amount of evaporated water 
that would be expected if the extra thermal heat (compared to 273~K) of the 300~K flow was 
simply converted into latent heat.

\subsubsection{Mechanisms warming the atmosphere}

\begin{figure*}
\begin{center}
\centerline{\includegraphics[scale=0.7]{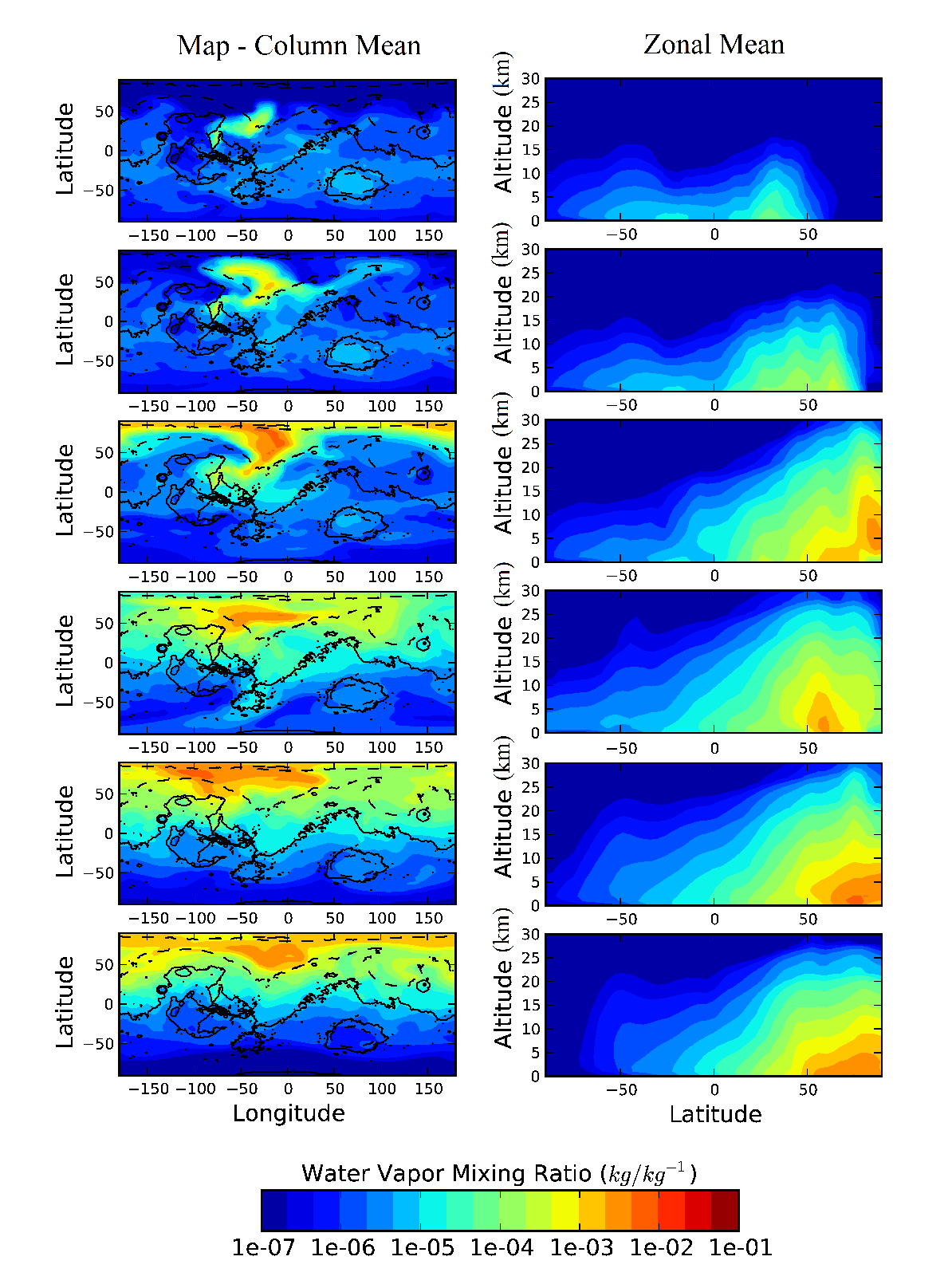}}
\caption{Time-lapse of the water vapor mixing ratio after (from the top to the bottom) 2.5/5/10/20/40/80 days. 
The left panels show the map of the water vapor distribution (column mean mixing ratio); 
the right panels show the corresponding zonal mean distribution (water vapor mixing ratio) as a function 
of latitude ($^{\circ}$N) and altitude (km).}
\label{vapor_map_lat}
\end{center}
\end{figure*}

As the water vapor starts to accumulate above the flow, the initially cold martian lower atmosphere 
soon reaches the water vapor saturation pressure. For instance, at 210 Kelvins, which is typically the 
mean surface temperature expected for a 0.2~bar atmosphere (Figure~\ref{control_simulation}), the water vapor 
saturation pressure is $\sim$~1.4 Pascals and the mass mixing ratio at saturation 
in a 0.2~bar atmosphere is thereby $\sim$ 7$\times$10$^{-5}$~kg/kg$^{-1}$.
This situation leads to the early condensation of the water vapor, latent heat release and thus to the warming of the atmosphere.
We identified this process as the dominant mechanism responsible for the warming of the atmosphere after an outflow event.

As the atmospheric temperatures increase, the capability of the atmosphere to retain water vapor also increases.
The mass mixing ratio at saturation, namely $Q_{\text{sat}}$, can be written as follows:
\begin{equation}
\label{qsat}
 Q_{\text{sat,H}_2\text{O}} = \frac{P_{\text{sat,H}_2\text{O}}}{P_{\text{CO}_2}+P_{\text{sat,H}_2\text{O}}},
~with~P_{\text{sat,H}_2\text{O}}(T)=P_{\text{ref}}~e^{\frac{L_{\text{v}}M_{\text{,H}_2\text{O}}}{R}(\frac{1}{T_{\text{ref}}}-\frac{1}{T})},
\end{equation}
with $P_{\text{sat,H}_2\text{O}}$ the water vapor saturation pressure and 
$P_{\text{CO}_2}$ the CO$_2$ partial pressure,
with $P_{\text{ref}}$ and $T_{\text{ref}}$ the pression/temperature of the triple point of water, 
respectively equal to 612~Pascals/273.16~Kelvins, 
$M_{\text{,H}_2\text{O}}$~$\sim$~1.8$\times$10$^{-2}$~kg~mol$^{-1}$ the molar mass of water, 
and $L_{v}$~$\sim$~2.26$\times10^6$~J~kg$^{-1}$ the latent heat of vaporization of liquid water. 
For low amounts of water, this relation simply becomes: 
\begin{equation}
\label{qsat_approx}
 Q_{\text{sat,H}_2\text{O}}(T) \sim \frac{P_{\text{ref}}}{P_{\text{CO}_2}}
~e^{\frac{L_{\text{v}}M_{\text{H}_2\text{O}}}{R}(\frac{1}{T_{\text{ref}}}-\frac{1}{T})}.
\end{equation}

Therefore, as the atmospheric temperatures increase, the atmosphere is also able to transport more and more water upwards.
Thus, as time goes on, the atmosphere becomes more and more warm and wet. As the atmospheric water vapor content increases, 
the absorption of the atmosphere in the infrared wavelength range (essentially due to the thermal emission of the warm outflow waters) 
increases and thus contributes to an additional warming of the atmosphere.

In total, during the warm phase (the first 500 days), the atmosphere (above the flow/lake) is directly 
warmed by the following processes (in decreasing order of importance): 1. the condensation of the water vapor produced 
by the warm flow ($\sim$~56~$\%$); 2. the sensible heat exchanged between the flow/lake and the lowest atmospheric layer ($\sim$~22~$\%$);  
3. the thermal infrared emission of the flow absorbed by the mixture of gaseous CO$_2$/H$_2$O ($\sim$~13~$\%$); 
and 4. the extra solar absorption resulting from the presence of water vapor excess, 
which has strong absorption lines in the solar domain ($\sim$~9~$\%$);. The atmospheric solar absorption is particularly important in this scenario, because 
we chose the outflow channel event to start at Ls~=~5$^{\circ}$ and thus to occur during the northern hemisphere spring and summer. 
Of course, all these processes reinforce and strengthen each other.

\begin{figure*}
\centering
\centerline{\includegraphics[scale=0.11]{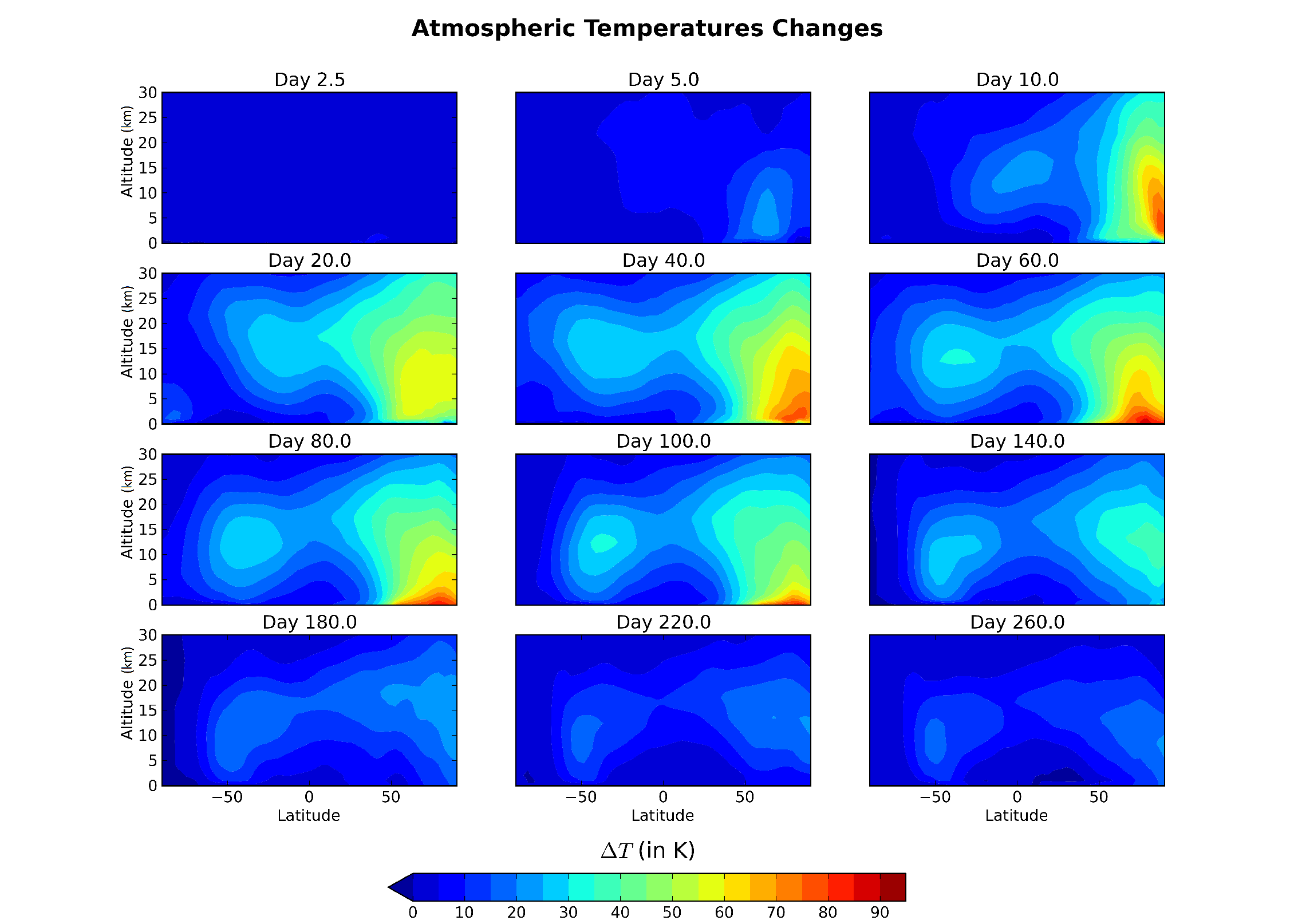}}
\caption{Time-lapse of the zonal mean cross-section atmospheric temperature difference between the reference simulation (with outflow) and 
the control simulation (without outflow) for the same surface pressure of 0.2~bar.}
\label{temp_plume}
\end{figure*}

Figure~\ref{vapor_map_lat} shows the spatial evolution of the water vapor atmospheric content. 
Initially, water vapor accumulates at low altitudes, in the regions where the liquid water flow is located. 
After a few days, the water vapor has reached much higher altitudes (up to $\sim$~30~km) through the 
aforementioned warming mechanisms and the convective adjustment scheme. 
Eventually, once the upper part of the atmosphere has become wet enough (typically after $\sim$~10 days in this scenario), 
the high altitude horizontal winds (around $\sim$~15~km) advect the water vapor into the neighbouring regions. 
After $\sim$~50 days, all the martian regions located above $\sim$ 50$^{\circ}N$ 
have become more or less wet, with a typical water vapor mean mass mixing ratio of 0.3$\%$. 

Similarly, the impact of H$_2$O condensation (and other additional warming sources) on atmospheric temperatures is 
shown in Figure~\ref{temp_plume}. After $\sim$~100 days, at the peak of the outflow channel event, 
the atmospheric temperatures in the lower atmosphere (0-5~km) almost reach 280~K, +90 Kelvins above the regular temperature 
(peak above the lake) as calculated in the control simulation; 
the atmospheric temperatures in the higher parts of the atmosphere typically extend up 
to 230~Kelvins (at 10~km) and to 170~Kelvins (at 25~km), which are respectively +50~K and +25~K above the temperatures prescribed by 
the control simulation.

\subsubsection{The mechanisms cooling the flow}

\begin{figure*}
\begin{center}
\centerline{\includegraphics[scale=0.75]{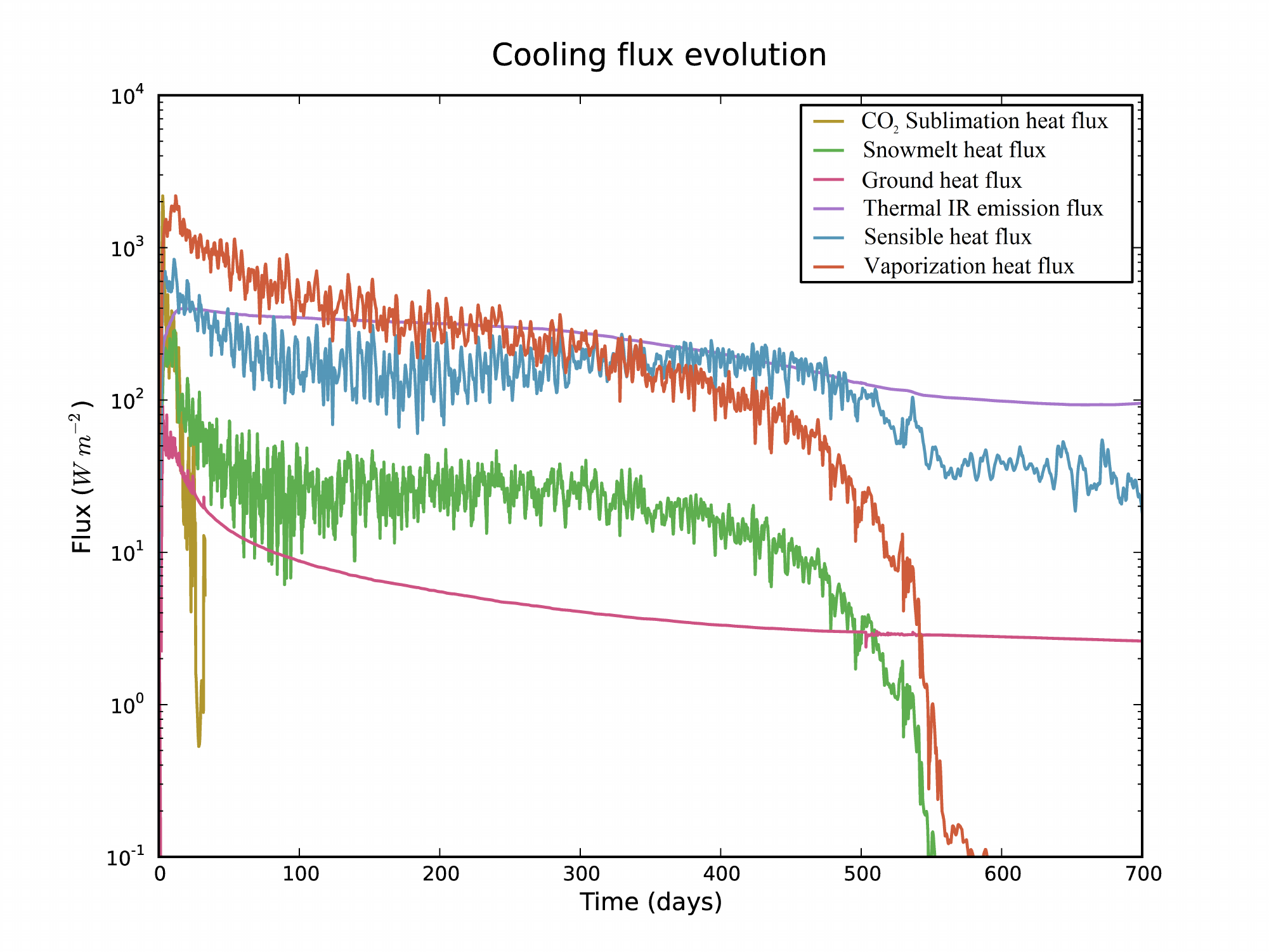}}
\caption{Surface cooling heat fluxes evolution averaged over all the Northern Plains lake grid cells in the 
P$_{\text{surf}}=0.2$~bar reference simulation. 
We note here that, depending on the nature and the intensity of 
a given outflow channel formation event, each of these fluxes can potentially become dominant.}
\label{cooling_surface_fluxes}
\end{center}
\end{figure*}

After $\sim$~500 days, which corresponds to the complete surface freezing of the outflow channel event water, 
the evaporation E produced by the stable lake (see Figure~\ref{evaporation_fig}) 
suddenly reduces (by almost 3 orders of magnitude). To a first order, we have in fact:
\begin{equation}
\label{qsat_approx2}
 E \propto Q_{\text{sat}}(T) \propto e^{-\frac{\alpha}{T}},
\end{equation}
with $\alpha=\frac{L_{\text{sub}}M_{\text{H}_2\text{O}}}{R}$ and $L_{\text{sub}}$ the latent heat of sublimation of water ice.
The evaporation rate E has thereby a strong dependence on temperature. This is why the drop 
in temperature associated with the surface freezing of the Northern Plains lake is responsible for 
the sudden decrease of evaporation visible in Figure~\ref{evaporation_fig} 
(also seen through the latent heat surface flux in Figure~\ref{cooling_surface_fluxes}).
This drop in evaporation defines the end of the 'warm phase', which includes the decrease of the water vapor content, 
the atmospheric temperatures and the precipitation events (see Figure~\ref{vapor_precip}).

There are several physical processes that are responsible for the cooling of the flow, leading to its solidification as ice.
Figure~\ref{cooling_surface_fluxes} shows the relative importance of the different thermal heat losses by the Northern Plains lake, 
from the beginning of the event to one martian year later.
For the first 500 days, the main cooling surface fluxes are the latent heat loss (420~W~m$^{-2}$, 43.3~$\%$), 
the sensible heat loss (190~W~m$^{-2}$, 19.6~$\%$), 
the radiative thermal emission loss (280~W~m$^{-2}$, 28.8~$\%$) 
and the ground conduction loss (8~W~m$^{-2}$, 0.8~$\%$).
Some other surface fluxes related to the CO$_2$ ice sublimation by the warm flow (13~W~m$^{-2}$, 1.3~$\%$) and 
the cooling of the lake by the melting of the falling snow (60~W~m$^{-2}$, 6.2~$\%$) also contribute to the cooling of the outflow waters. 
In total, the average cooling flux of the outflow waters for the warm phase (first 500 days) is $\sim$~970~W~m$^{-2}$.

\begin{figure*}
\begin{center}
\centerline{\includegraphics[scale=0.15]{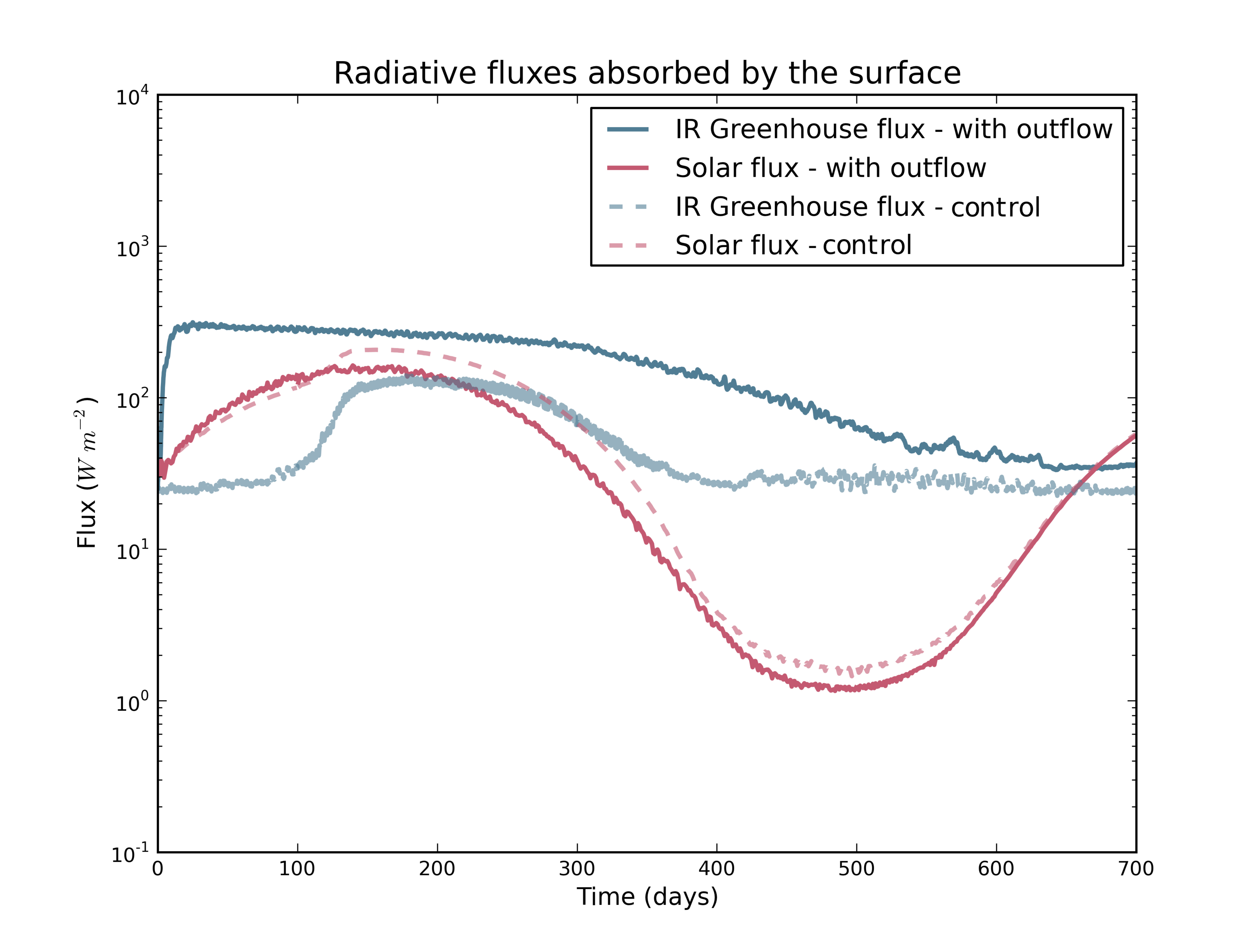}}
\caption{Evolution of the radiative fluxes absorbed by the surface and  averaged over all the Northern Plains lake grid cells. 
Solid lines refer to the solar flux (in red) and the thermal infrared (in blue) for the reference simulation. 
Dashed lines correspond to the control simulations.
For better visibility, we filtered diurnal waves from the absorbed solar fluxes using a 1 day running average.}
\label{warming_surface_fluxes}
\end{center}
\end{figure*}

For large outflow channel formation events like the one described in this section, 
the sublimation of the seasonal carbon dioxide ice deposit represents a small fraction of the heat loss. Nonetheless, 
smaller outflow channel events (5$\times$~$10^3$~km$^3$ for example \citep{Andr:07}) flowing on the Northern Plains slopes may be deeply affected by the energy gap 
required to sublimate the CO$_2$ ice seasonal deposit. For a 0.2~bar atmosphere, the control simulations show, for example, that the CO$_2$ ice 
seasonal deposit reaches a yearly average of $\sim$~300~kg~m$^{-2}$ from the North Pole down to 30$^\circ$N latitudes.

Two radiative processes may counteract the cooling of the flow: 1) the absorption of solar radiation and 2) the greenhouse effects 
(of the atmosphere and of the clouds).

1. We chose in this scenario to start the outflow channel event at Ls~=~5$^{\circ}$ in order to maximize the role of solar absorption.
The peak of the event (between $\sim$0-300 days, Ls~$\sim$~5-165$^{\circ}$) was therefore chosen to overlap 
with the peak of insolation in the Northern hemisphere, which is a maximum of $\sim$~170 days after the event (Ls~=~90$^{\circ}$).
There are three factors that need to be taken into account in the solar absorption processes: absorption by water vapor, 
albedo changes and clouds. For this reference simulation, compared to the control simulation, these three effects more or less compensate at the location 
of the flow. The increase of the solar absorption due to the low albedo of liquid water (0.07 compared to 0.2 for the bare ground and 
0.5 for the remaining CO$_2$ ice seasonal cover) and due to the absorption by water vapor are more or less balanced by the 
reflection of the cloud cover, which can reach on average a coverage of 80~$\%$ during the first 500 days above the lake (Figure~\ref{cloud_coverage}). 
Most of these water clouds are located at low altitude (Figure~\ref{cloud_coverage}). 
During the warm phase, the lake absorbs a solar flux of $\sim$~67~W~m$^{-2}$ ($\sim$~16~W~m$^{-2}$ less than the control run, 
see Figure~\ref{warming_surface_fluxes}) and the atmosphere 
(essentially the troposphere) $\sim$~20~W~m$^{-2}$ ($\sim$~12~W~m$^{-2}$ more than the control run). 
This corresponds to an average absorption of 65~$\%$ of the available incoming solar flux ($\sim$~135.6~W~m$^{-2}$ for the first 500 days).

\begin{figure*}
\begin{center}
\centerline{\includegraphics[scale=0.8]{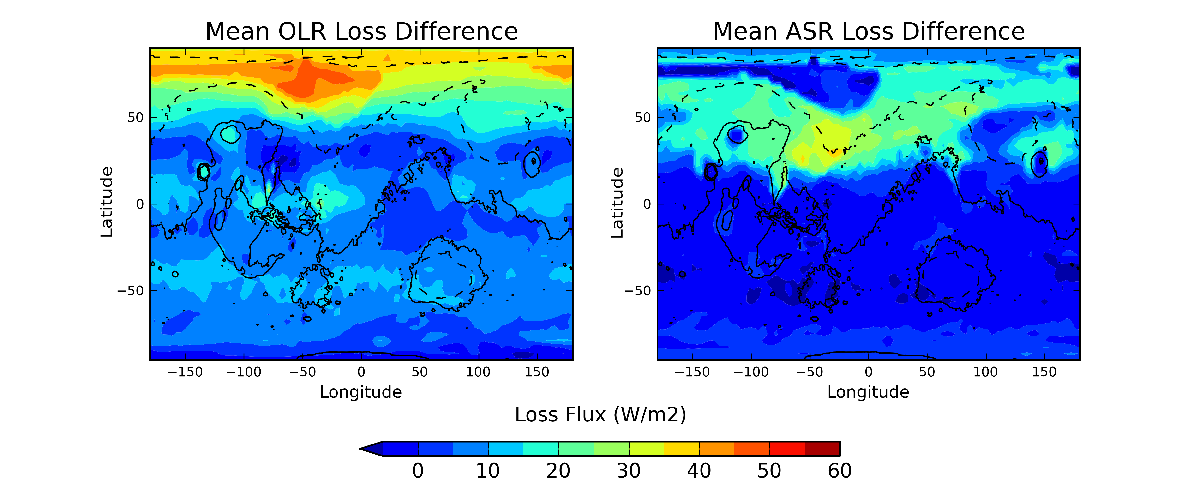}}
\caption{Mean Outgoing Longwave Radiation (OLR, left) and Absorbed Solar Radiation (ASR, right) loss during the warm phase,
for the reference simulation, and relative to the control simulation performed for the same surface pressure.}
\label{asr-olr_map}
\end{center}
\end{figure*}

\begin{figure*}
\begin{center}
\centerline{\includegraphics[scale=0.65]{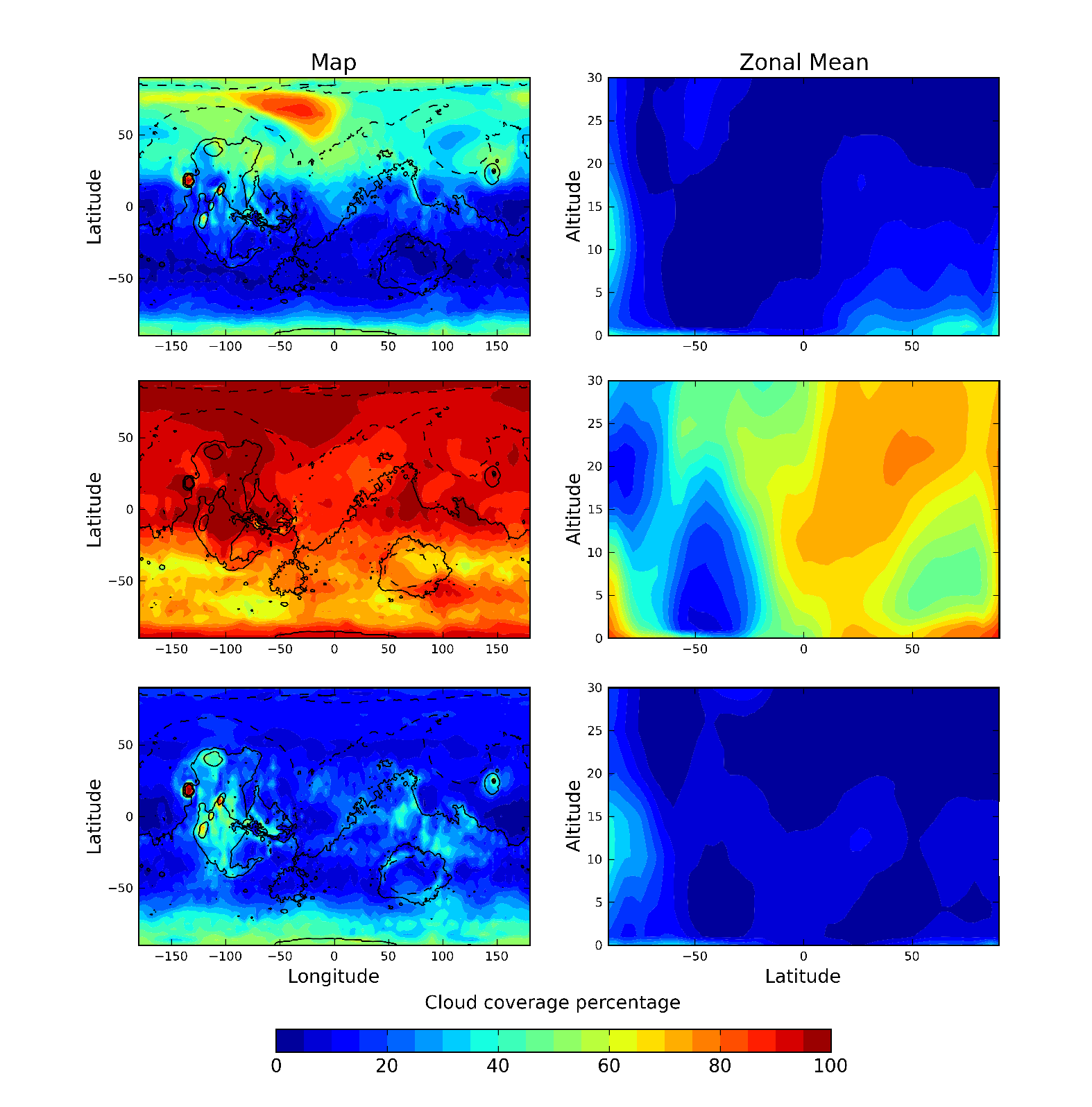}}
\caption{One year average of the cloud coverage following the outflow event. The first row corresponds to the 
map and the zonal mean cross-section of the reference simulation. In the second row, precipitation was removed (see section~\ref{urata_test}).
The third row is for the control simulation.}
\label{cloud_coverage}
\end{center}
\end{figure*}

2. The downward thermal infrared emission from the atmosphere and the clouds is 
the dominant warming flux (see Figure~\ref{warming_surface_fluxes}). On average, during the warm phase, 
this greenhouse effect brings $\sim$~210~W~m$^{-2}$ to the lake (+~150~W~m$^{-2}$ more than the control run). 
The main source of thermal infrared emission surface heating comes from the gaseous atmosphere itself, 
which can reach up to $\sim$~280~K (above the lake) for the first 5~km, at the peak of the event.

In total, both solar and infrared heating counterbalance only $\sim$~30$\%$ of the cooling of the flow, and are thus 
unable to sustain the perturbation generated by the ouflow channel. 
We note here that the radiative effect of H$_2$O clouds during the warm phase 
is approximately neutral or at least very limited (only +7~W~m$^{-2}$) above the Northern Plains lake, 
with +23~W~m$^{-2}$ of greenhouse warming and -17~W~m$^{-2}$ due to the reflection of the sunlight.

\subsubsection{The mechanisms cooling the atmosphere}

One of the main results of our work is that outflow channel events 
are not able to sustain warm conditions. We present here the two processes 
that act efficiently together to cool down the atmosphere after outflow events.

1. In the time following catastrophic outflow channel events like the one described in this section, 
the atmosphere above the flow warms very quickly. In our reference simulation, 10 days after 
the beginning of the event, the temperature in the lower atmosphere (0-5~km) above the lake 
increases by almost 90~Kelvins. During the first 500 days after the event, because of this significant warming, 
the flow and the atmosphere just above it contribute 
to an extra thermal infrared emission loss to space of 38~W~m$^{-2}$ compared to the control simulation.
Yet the amount of energy lost by the lake and the atmosphere above represent only $\sim$~11$\%$ of the extra total cooling to space. 
Figure~\ref{temp_plume} shows that, as the atmosphere gets warmer in the regions of the flow, high altitude winds around 
$\sim$~15~km advect the heat to the neighbouring areas (in particular into the Northern Plains).
This increases the surface of the emissions and therefore strengthens the cooling.

Figure~\ref{asr-olr_map} (left) shows the regions of the planet responsible 
for the extra thermal emission to space. Globally, during the warm phase (the first 500 days), 
the planet loses $\sim$~10~W~m$^{-2}$. One third of the emissions are due to the regions of latitude $>50^{\circ}$N.
During the warm phase, the most important mechanism of cooling is the thermal infrared emission, 
enhanced by the advection processes.

2. Interestingly, another important cooling mechanism is the decrease of solar absorption due to the increase of surface albedo that 
follows the outflow channel event. In fact, the precipitation caused by the event, 
essentially in the form of snowfall (see Figure~\ref{vapor_precip}), leaves ice (see Figures~\ref{season_deposit} and \ref{asr-olr_map}) 
over an area of $\sim$~30~$\times$~10$^6$~km$^2$ that reflect an important part of the sunlight ($\sim$~21.5~W~m$^{-2}$).
In total, during the warm phase and compared to the control simulation, 
the decrease of solar absorption contributes to a global equivalent extra cooling of $\sim$~4.5~W~m$^{-2}$, which represents 
half of the infrared emission loss to space.

The large amount of water vapor released after the outflow channel event condenses very quickly in the atmosphere, forming clouds that are 
mostly located in the area of the flow and of the resulting lake (see Figure~\ref{cloud_coverage}). In total, for the reference simulation, 
the clouds have a slight positive effect of +1.3~W~m$^{-2}$ (+~2.3~W~m$^{-2}$ of greenhouse effect and 
-~1.0~W~m$^{-2}$ of solar reflection).

\subsubsection{Consequences on the water cycle and the precipitation}

\begin{figure*}
\centering
\centerline{\includegraphics[scale=0.6]{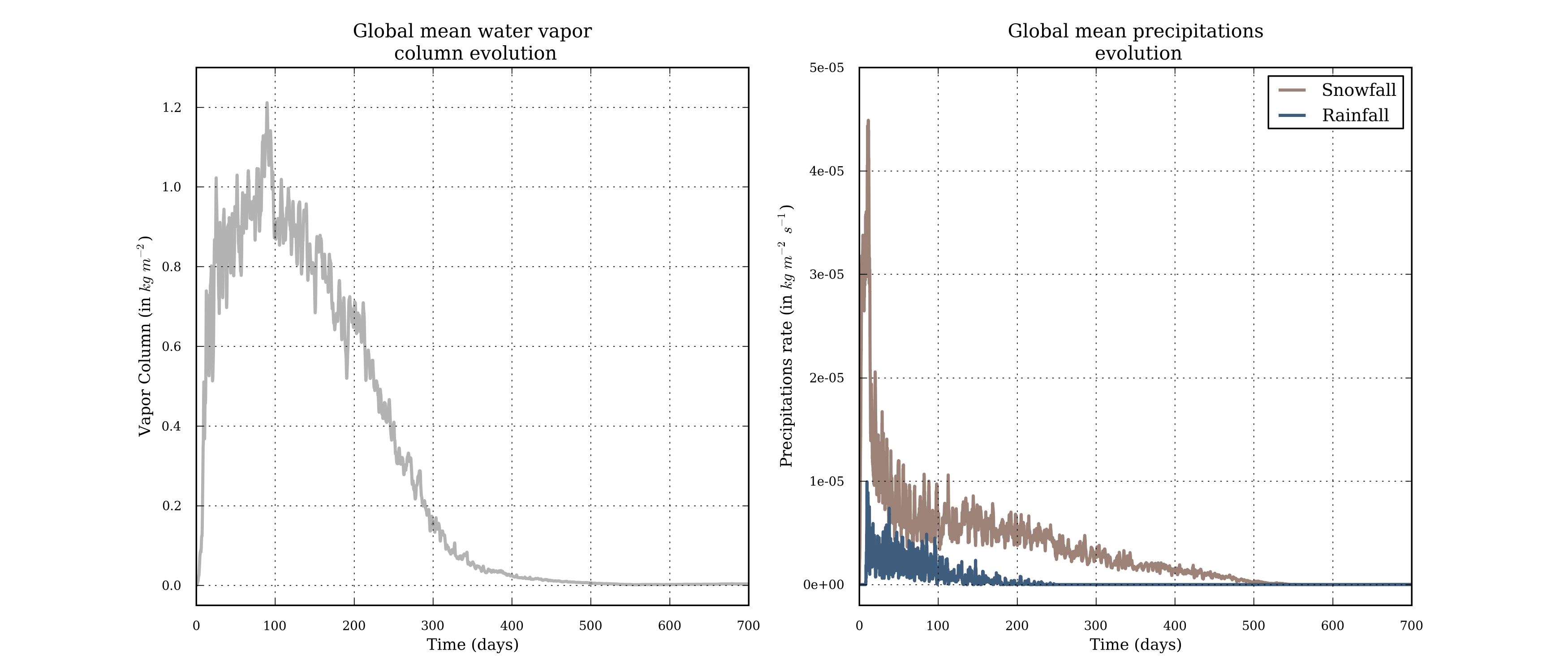}}
\caption{Evolution of the global mean water vapor column (left) and the precipitation (right), 
during the year following the outflow channel reference event.}
\label{vapor_precip}
\end{figure*}

The maximum total amount of water vapor that is carried by the atmosphere during the event (GEL of 1.2~mm at the peak) 
remains limited by comparison to the cumulative total amount of precipitable water generated (GEL of 230~mm). 
It represents only $\sim$~0.5$\%$ of the cumulative evaporated water vapor 
produced by the entire ouflow channel event during the first 500 days. 
Figure~\ref{vapor_precip} (left) shows the global mean atmospheric 
water vapor content (column mass in kg~m$^{-2}$, and also GEL in mm). It peaks at $\sim$~100 days and 
considerably decreases from $\sim$~200 days to $\sim$500 days.

The fact that the atmosphere is not able to accumulate more than $\sim$~1.2~kg~m$^{-2}$ (globally) and $\sim$~50~kg~m$^{-2}$ (locally, 
just above the warm lake) has one main consequence: the atmosphere does not manage to carry enough water vapor far enough from the lake to create 
precipitation in regions of interest (West Echus Chasma Plateau in particular). 
The typical lifetime of the atmospheric water vapor is in fact $\sim$~0.5 days 

\begin{figure*}
\begin{center}
\centerline{\includegraphics[scale=0.5]{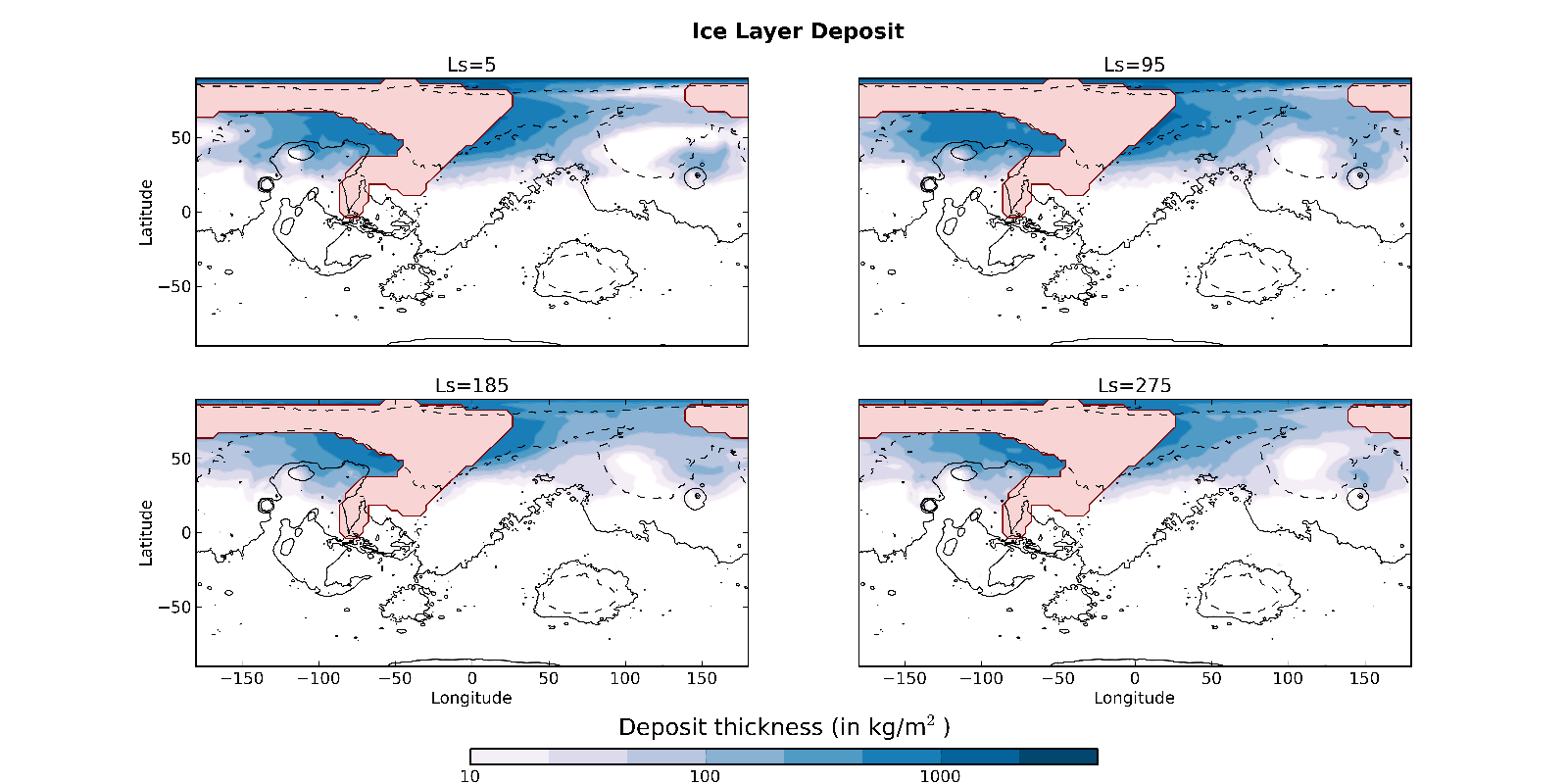}}
\caption{Ice deposit field obtained 1 martian year after the beginning of the event, 
for four different start dates (Ls~=~5$^{\circ}$, 95$^{\circ}$, 185$^{\circ}$, 275$^{\circ}$). 
The first figure (on the top left corner) corresponds to the reference simulation and starts at Ls~=~5$^{\circ}$. 
We use the pink color for the regions where the flow passed through on its way to the lake.}
\label{season_deposit}
\end{center}
\end{figure*}

Rainfall, which represents a very small fraction ($\sim$~10~$\%$) of the precipitation (Figure~\ref{vapor_precip}), occurs only above the 
Northern Plain lake, because this is the only location of Mars where atmospheric temperatures exceed (up to 10~km) the temperature of the triple point.
Outside the lake, the only mechanism of precipitation is snowfall. Approximately 50~$\%$ of the snow falls back directly on the flow/lake. 
The rest of the precipitation (the 50~$\%$ remaining) is essentially confined in the northern regions.
Figure~\ref{season_deposit} shows the map of the deposited ice field (generated by precipitation) after a simulation of one martian year.
The fraction of this ice that is melted after an outflow event is very limited (see Figure~\ref{psurf_max_melted}), 
because 1) most of the thermal perturbation has been dissipated by 
advection/cooling to space processes after $\sim$~200 days, 
2) the remaining water vapour abundance after these 200 days is too low to trigger a significant 
greenhouse warming (as found by \citealt{Kite:11a}) 
and 3) the ice field itself raises the albedo of the surface and thus acts as a very efficient climatic cooling agent.

In summary, the short-term climatic impact of outflow channel formation events seems very limited. 
For a 0.2~bar atmosphere, an outflow channel event of 10$^6$~km$^3$/300~K leads to the formation of a lake (located in the 
Northern Plains main topographic depression) that triggers a warm period that lasts for 
$\sim$~500 days, which coincides approximately with the complete surface freezing of the water in the lake.
Such events leave globally $\sim$~6.5~$\times$~10$^3$~km$^3$ of water ice/snow (0.65$\%$ of the initial outflow reservoir) 
and are able to melt $\sim$~80~km$^3$ (0.008$\%$ of the initial reservoir; 1$\%$ of the deposited precipitation).
Because the outflow events do not manage to warm the atmosphere enough, water vapour stays confined to the regions neighbouring the lake 
(essentially in the Northern Plains) and therefore precipitation (mostly snowfall) and melting only occur in the lowland regions.

The long-term climatic impact of the ice-covered lake is discussed in the next section.

\subsection{The Cold Phase}

After 500 martian days, the surface of the Northern Plains lake is completely covered by ice. 
 Temperatures, water vapor content and precipitation all decrease. Because the area of high albedo ice deposits is larger 
than in the control simulations, the mean surface temperatures extend even lower than before the ouflow event 
(-2~K for the global annual surface temperatures of the 0.2~bar reference simulation, and compared to the control simulation).

Using the extrapolation scheme presented in section~\ref{param_outflow}, 
we estimated that the released water was completely frozen after $\sim$~4~$\times$~10$^3$ martian years.
This corresponds to the full solidification of the water to ice 
at the location of the main Northern Plains topographic depression (which is the deepest point of the lake). After $\sim$~500~years, 
more than 70~$\%$ of the lake (in area) is frozen, from the surface to the top of the regolith. We note that the ground thermal flux 
\citep{Clif:01} during the Late Hesperian era was one order of magnitude too low (at best) 
to be able to increase the lifetime of the $\sim$~500~m deep lake.

In our simulations, $\sim$~10~years after the beginning of the lake-forming event, the mean ice thickness over the lake is $\sim$~25~meters. 
The annual mean conduction heat flux for this ice thickness is $\sim$~10~W~m$^{-2}$. 
The annual mean solar/IR fluxes absorbed by the ice are $\sim$~53/57~W~m$^{-2}$ by comparison, 
110~W~m$^{-2}$ in total. Under these conditions, the thermal conduction flux represents less than 10~$\%$ of the total heat flux 
received by the surface at the location of the lake.
Moreover, because the temperature profile oscillates annually, 
in the first 5 meters (typically) of the ice cover, from positive values (summer season) to negative 
values (winter season), the heat conduction from the liquid water to the surface is mainly returned during the winter seasons. Yet, the water cycle in this 
cold phase is essentially controlled by the summer seasons, because sublimation rates are several orders of magnitude higher than during the winter seasons 
(see section~\ref{param_outflow} for discussion).
Thus, after a few years (typically around 10), the climatic effect of the lake becomes, to a first order, the same 
as simply placing a comparable-sized body of ice in the Northern Plains. During these 10 years, 
ice transportation/water vapor cycle/precipitation is very limited by comparison to the warm phase and do not play 
any significant role in the ice field position. 

Within the lifetime of the liquid water lake, the ice field position evolution is completely controlled by the $\sim$~4~$\times$~10$^3$~(-10)~years 
of the water cycle forced by the sublimation of the large body of non-stable ice.

Each year, during Northern summer, $\sim$~20~mm~year$^{-1}$ of lake ice sublimes to condense elsewhere and 
approximately 30~$\%$ of it is transported away from the lake. Progressively, the water vapor produced during the summers 
migrates southward and - through the mechanism of adiabatic cooling - condenses on the regions of high altitudes and low latitudes.
The lifetime of the frozen lake predicted by our simulations is $\sim$~7~$\times$~10$^4$~martian years. 

\begin{figure*}
\begin{center}
\centerline{\includegraphics[scale=0.12]{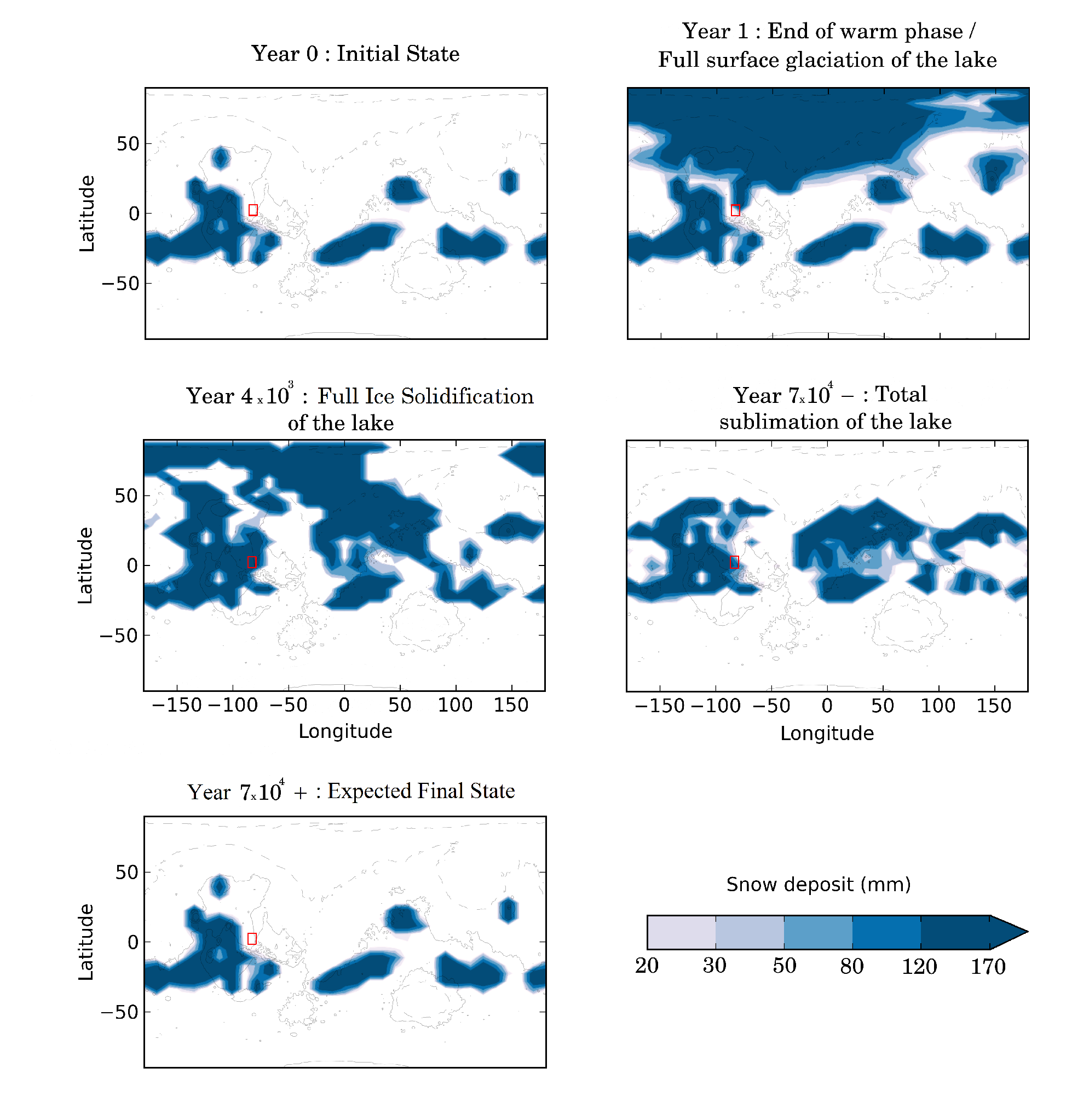}}
\caption{Ice deposit field after 0~/~1~/~4~$\times$~10$^3$~/~7~$\times$~10$^4$ martian years, 
corresponding to the main phases following the outflow channel reference event.
The red rectangle corresponds to the West Echus Chasma Plateau area.}
\label{ice_chrono}
\end{center}
\end{figure*}

The evolution of the ice field through the phases that follow the outflow channel reference event are shown in Figure~\ref{ice_chrono}.
After $\sim$~10$^5$ martian years, the outflow channel water is located more or less exclusively in the highland regions.
During this cold phase ($\sim$~10$^5$ martian years), some ice appears stable in the region of West Echus Chasma Plateau, 
due to the uninterrupted supply of ice coming from the northern parts of the planet. 
This snow deposit is produced by the adiabatic cooling of the ascending air masses 
that provoke the condensation of the water vapor initially generated by the sublimation of the Northern Plains ice field. 

Some water ice is also transported to the drainage regions of Alba Patera, Hecates Tholus and Ceraunius Tholus but 
this might not be a critical factor since our model already predicts that ice deposits should be stable 
in these regions (\citealt{Word:13}, Figure 2 – this work) and therefore available for either seasonal snowmelt or ground melting. 

In spite of this, because the global surface albedo is increased during that period, global temperatures are much lower than before the outflow event, 
making snowmelt difficult.

We note here that we did not take into account the flow of the ice on the Northern Plains slopes. This could significantly increase the lifetime of the 
lake located in the main topographical depression and thus the lifespan of the snow deposited in non-stable locations (in particular in West Echus 
Chasma Plateau area). 
However, at these temperatures and over these timescales, ice is unlikely to flow significantly \citep{Fast:12,Fast:14,Fast:15}.
In addition, we did not take into account the formation 
of a possible lag deposit \citep{Kres:02jgr,Moug:12} which could have decreased the sublimation rate of the ice. 
Both of these factors, however, appear to have minimal effects on the general processes.

\subsubsection{Influence of obliquity}

Orbital spin-axis obliquity is a very important factor in the duration and the characteristics of the cold phase, because 
it controls the latitudinal distribution of the solar flux and thus the sublimation processes.
We performed two simulations of the reference outflow channel event, at obliquities of 25~$^{\circ}$ and 65~$^{\circ}$, 
to complement the 45~$^{\circ}$ obliquity case presented initially.

In the low obliquity simulation, the sublimated ice migrates slowly toward the coldest points of the planet: 
the South pole and the North pole (in agreement with \cite{Word:13}, Figure~4).
The water present in the northern part of the lake is stable in the long term.
In this situation, ice never accumulates in the region of West Echus Chasma.

In the high obliquity simulation, the water cycle is much more intense 
because the peak of insolation at high latitudes is higher.
Approximately $\sim$~55~mm of the sublimated northern lake ice migrates southward each year. 
The lifetime of the lake is thereby lowered to $\sim$~9~$\times$~10$^3$ martian years. 
For the same reasons as that in the reference simulation, a thick ice deposit is present in the region of West Echus Chasma Plateau. 
Yet, its duration, $\sim$~10$^4$ years, is almost 10 times less than in the reference simulation, 
more or less coincident with the lifetime of its supply (the frozen lake).

As a result, the lifetime of the ice located in West Echus Chasma area seems to be favored at obliquity $\sim$~45~$^{\circ}$.

\section{The effect of surface pressure}

For many reasons (see discussion in section~\ref{LH_climate}), the atmospheric pressure during the Late Hesperian epoch 
is not well constrained. We explore in this section the role of surface pressure on the 
climatic impact of outflow channels.

For this, we performed five different simulations of the same outflow channel event 
(10$^6$~km$^3$, 300~K water released at 1~km$^3$~s$^{-1}$ in Echus Chasma) 
for five different surface pressures (40~mbar, 80~mbar, 0.2~bar (the reference simulation), 0.5~bar and 1~bar).

\subsection{Warm Phase}

Atmospheric pressure is one of the key factors that control the efficiency at which the warming of the atmosphere 
and the transport of water occur during the warm phase, as pointed out by \citet{Kite:11a}.

1. The evaporation rate: Combining equations~\ref{evap} and \ref{qsat_approx} for low amounts of water vapor,
the evaporation rate $E$ can be written:
\begin{equation}
\label{evap_psurf}
E~=~\frac{C_d V_{1} P_{\text{ref}} M_{\text{CO}_2}}{R~T_{1}}~e^{\frac{L_{\text{v}}M_{\text{H}_2\text{O}}}{R}(\frac{1}{T_{\text{ref}}}-\frac{1}{T_{\text{surf}}})}.
\end{equation}
Hence, the evaporation rate does not (directly) depend on the surface pressure and is 
mostly controlled by the temperature T$_{\text{surf}}$ of the flow/lake. 
To first order (and this is confirmed by our simulations), the wind velocity $V_1$ and the atmospheric 
temperatures $T_{1}$ do not differ sufficiently from one atmospheric pressure to 
another to play a major role on the rate of evaporation.

2. The warming rate: The volumetric heat capacity of the atmosphere increases linearly with the volumetric mass density and thus the atmospheric pressure.
For example, it takes approximately $\frac{1.0}{0.040}~=~25~\times$ more energy to warm a 1~bar atmosphere than a 40~mbar one.

When the outflow channel event occurs, the rate of warming of the atmosphere (in K/s) is 
roughly proportional to the evaporation rate (which is the main source of 
heating) and inversely proportional to the volumetric heat capacity of the atmosphere. 
In our simulations, it takes $\sim$~10/40 martian days - respectively for the 40~mbar/1~bar case - 
for the atmospheric temperatures at 10~km to reach a plateau at 
250~K/220~K, which correspond to a +80~K/+30~K temperature increase (for initial temperatures equal to 170~K/190~K). 
This corresponds approximately to a factor of 10 in heating efficiency for these two endmember situations. 
The difference between the factor of 25 predicted and the factor of 10 obtained in our simulations is mostly due to two processes: 
advection and thermal emission to space.

The same two processes limit the growth of atmospheric temperatures. 
First, the advection tends to dilute the heat perturbation horizontally. 
In the 1~bar case, this is the dominant process for example.
Second, the thermal emission to space acts as a very efficient negative feedback. 
This is, in fact, the first limiting process in the 40~mbar case.

\begin{figure*}
\begin{center}
\centerline{\includegraphics[scale=0.8]{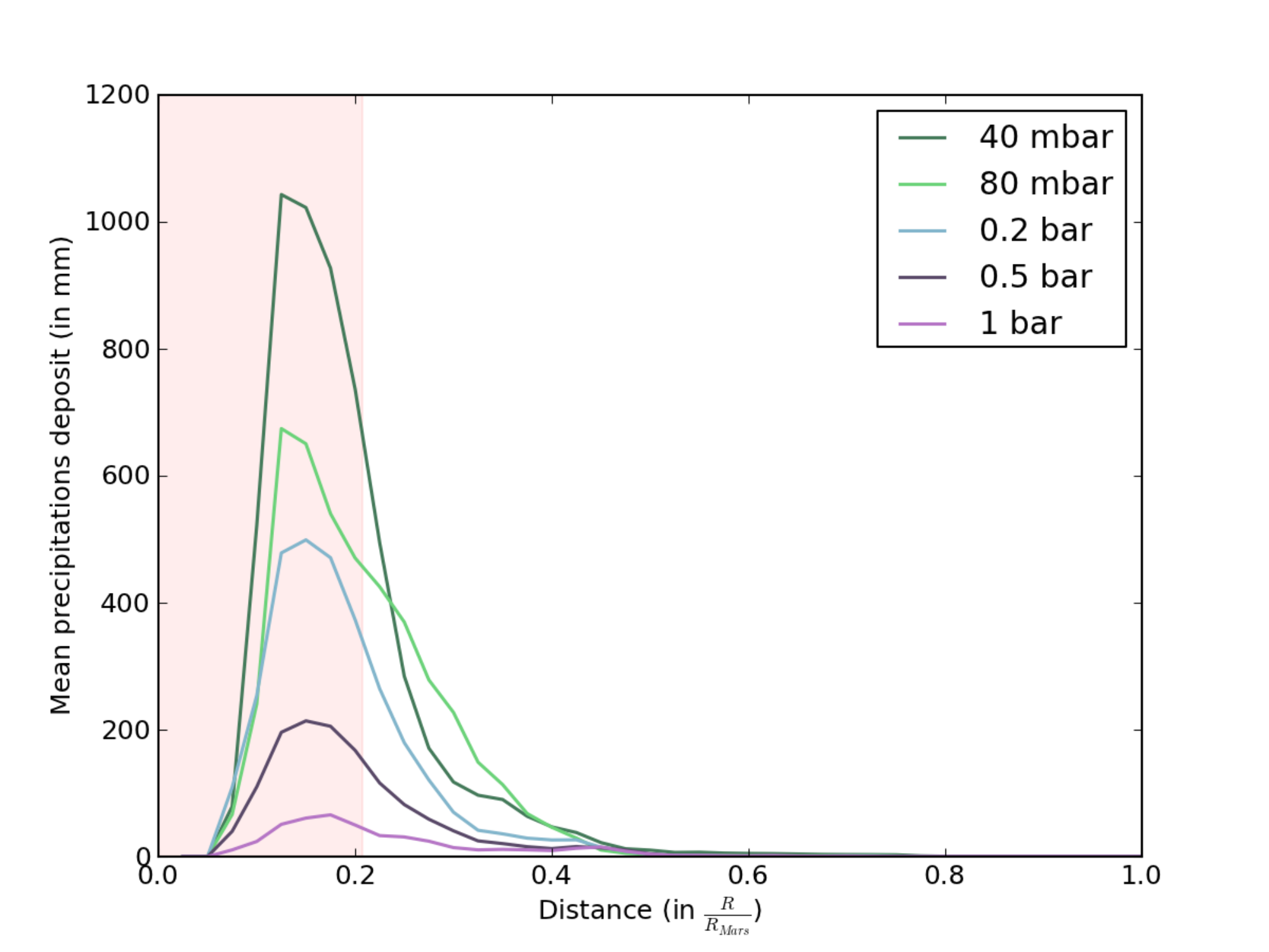}}
\caption{Radial precipitation distribution around the center of the lake (70$^{\circ}$N,-30$^{\circ}$E), 
averaged for the first 500 days following the outflow channel event, and for 5 different atmospheric pressures. 
The precipitation falling back on the lake/flow was removed from the plot. 
Because the lake is not circular, we used the pink color 
to represent the maximum radial extent of the lake.}
\label{radial_distrib}
\end{center}
\end{figure*}

The capability of an atmosphere to maintain high temperatures from the surface (where evaporation occurs) to the altitude where advection occurs 
is in fact the most important factor in the ability to transport water vapor globally and produce precipitation far from the region of evaporation.
The warmer the atmospheric column above the lake is, the more water vapor will be possibly lifted and then transported globally 
by the high altitude winds.

Thin atmospheres (such as the 40~mbar) warm efficiently above the region of the flow, 
allowing the formation of a persistent water vapor plume that can transport (through advection) water vapor far from the flow/lake. 
In contrast, thick atmospheres (such as the 1~bar case) ironically do not manage to transport water efficiently because of the advection itself. 
The advection prevents the atmospheric temperatures above the lake from building up and thus the water vapor from accumulating.
This limits the transport of water vapor and favors local precipitation.
This is summarized by Figure~\ref{radial_distrib} that shows 
the radial mean distribution (centered above the Northern Plains lake) of precipitation for 
the entire warm phase (first 500 days). Our experiments show that thin atmospheres are able to transport much more water 
and for much longer distances than thick ones.

\begin{figure*}
\begin{center}
\centerline{\includegraphics[scale=0.7]{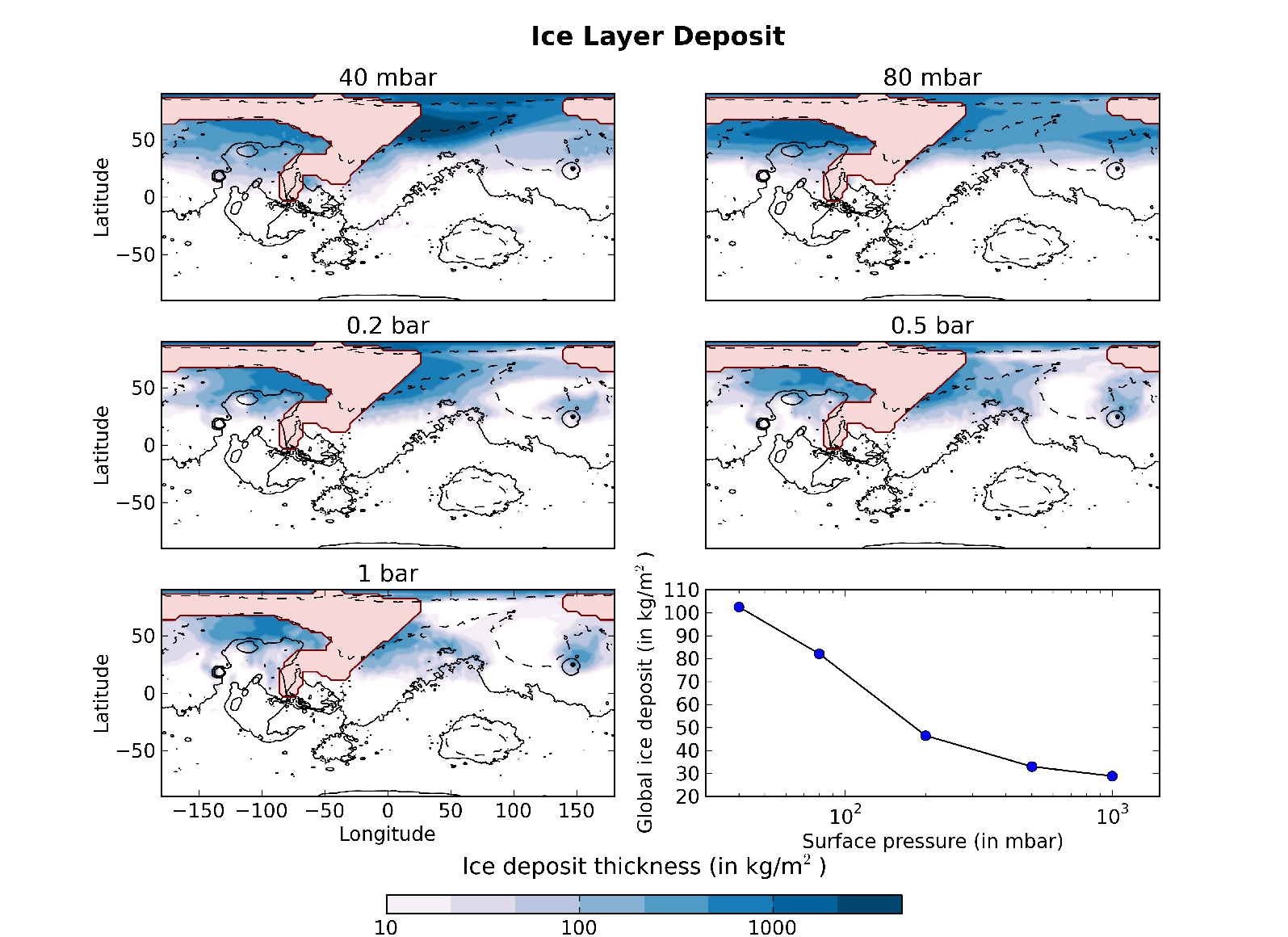}}
\caption{Final Ice layer deposit map (in kg~m$^{-2}$) after 1 martian year of simulations, 
for five different surface pressures (40~mbar, 80~mbar, 0.2~bar, 0.5~bar and 1~bar). 
The pink color denotes the regions where the flow passed through on the way to the lake.}
\label{psurf_ice_deposit}
\end{center}
\end{figure*}

We compare in Figure~\ref{psurf_ice_deposit} the spatial distribution of the precipitation (only snowfall, because rainfall 
occurs only above the lake) for the different atmospheric pressures. Whatever the surface pressure considered, the precipitation 
stays confined to the Northern Plains.

\begin{figure*}
\begin{center}
\centerline{\includegraphics[scale=0.7]{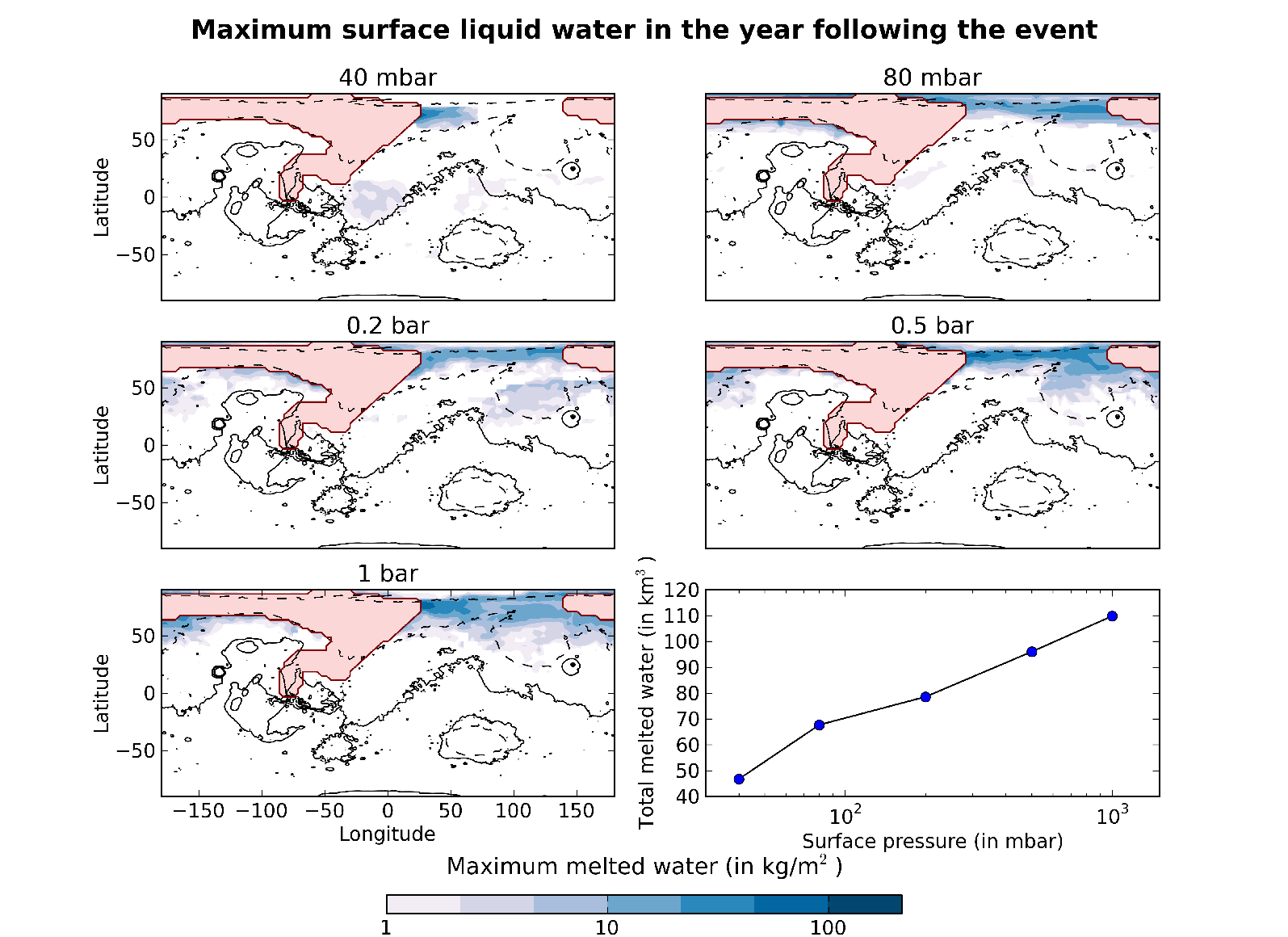}}
\caption{Maximum surface liquid water after 1 year of simulation and for 
five different surface pressures (40~mbar, 80~mbar, 0.2~bar, 0.5~bar and 1~bar). 
The pink color denotes the regions where the flow passed through on the way to the lake.}
\label{psurf_max_melted}
\end{center}
\end{figure*}

Another important aspect concerns the role of atmospheric pressure on the ability to melt the ice initially present / transported by the 
outflow event itself. 
Thin atmospheres, while able to reach temperatures in excess of 273~K above the flow, are not able to raise global temperatures significantly. 
First, the relaxation timescale of the temperature field is very low in such atmospheres because of the weak infrared absorption of 
the atmosphere.
Second, outflow channel events under thin atmospheres generate a very large ice cover that reflects sunlight efficiently.
As a result, an outflow channel of 10$^6$~km$^3$ that occurs under a 40~mbar atmosphere, 
leaves globally $\sim$~1.5~$\times$~10$^4$~km$^3$ of water ice/snow (1.5$\%$) and is able to melt only $\sim$~50~km$^3$ (0.005$\%$).

Thick atmospheres are initially warmer than thin atmospheres (+~30~K between the 1~bar and 40~mbar atmospheres). 
They also have a much more efficient infrared absorption and thus 
are better candidates to melt the deposited ice field. 
For example, an outflow channel of 10$^6$~km$^3$ that occurs under a 1~bar atmosphere, 
leaves globally $\sim$~4~$\times$~10$^3$~km$^3$ of water ice/snow (0.4$\%$) and is able to melt $\sim$~110~km$^3$ (0.011$\%$).

Nonetheless, this melting occurs only in the Northern Plains, in the close vicinity of the lake, because such thick atmospheres do not transport 
much ice anywhere on the planet in any case. In addition, ice albedo feedback (which is yet lower for thicker atmospheres) 
and the high volumetric heat capacity (lower heat perturbation) 
of such atmospheres contribute to lower the possibility of reaching melting temperatures.

Whatever the value of the surface pressure, the ability of the atmosphere to produce liquid water from melting is very limited.

\subsection{Cold Phase}

The water cycle during the cold phase is, in contrast, more intense for thick atmospheres than for thin ones. The sublimations rates are higher 
because global temperatures (and also summer temperatures) are also higher.
At the end of the warm phase, the mean global temperatures for the 40~mbar/1~bar simulations are 
respectively $\sim$~193~K (3.5~K lower than the control simulation)  and $\sim$~226~K (1~K lower than the control simulation). This 
difference is due to the increased ice cover following the outflow event.

In the 1~bar simulation (thick case), the lifetime of the frozen lake is $\sim$~5~$\times$~10$^4$ martian years, slightly lower than in the 
reference simulation. The climatic response during the cold phase behaves more or less in the same manner as in the reference 0.2~bar simulation.

In the 40~mbar simulation (thin case) however, because the water cycle is too weak (sublimation rate of the lake of 2mm/year; 
lifetime of the frozen lake $\sim$~2~$\times$~10$^5$ years), the southward flux of the atmospheric water ice 
is not high enough to allow the presence of stable ice in the area of the West Echus Chasma Plateau.

More generally, atmospheres with pressure higher than 80~mbar seem necessary to produce ice deposits in the region of West Echus Chasma Plateau.

\section{Extreme parameterizations}

In this section, we study several scenarios that may deeply affect the climatic impact of outflow events: 
1. the intensity of the event and 2. the effect of clouds and precipitation.

\subsection{Intensity of the event}

Because outflow channel events such as the one presented in Section~\ref{reference_results} fail to produce rainfall/transient warming, 
it is tempting to explore even more extreme parameterizations of the outflow events.

\subsubsection{Temperature of the flow}
\label{temp_flow}

The temperature of the groundwater released during outflow events is not well constrained (see section~\ref{outflow_description}). 
Hence, we used the temperature of the flow as a tuning parameter to explore the sensitivity of our results to the intensity of the outflow event.
We performed three simulations of the same outflow event (10$^6$~km$^3$, released in Echus Chasma) for three different groundwater temperatures: 
280~K, 300~K (reference simulation) and 320~K.

As expected, the warmer the water, the more intense the climatic effect becomes. For example, at the peak of the warm phase, 
the 320~K event is able to carry approximately 8~$\times$ more water vapor than in the reference simulation because atmospheric warming processes 
are amplified by the temperature (evaporation/condensation cycle, IR emission of the flow, ...). 
Consequently, 25~$\%$ of the precipitation following the 320~K event is 
rainfall (respectively 10~$\%$/~0~$\%$ for the reference/280~K simulations). Yet, rainfall still occurs exclusively above the lake (70~$\%$) or 
in the northern lowlands of Mars (30~$\%$). Snow precipitation also remains confined to the Northern Plains 
down to 15~$^{\circ}$N (25/40~$^{\circ}$N  for the 300~K/280~K simulations).

\begin{figure*}
\begin{center}
\centerline{\includegraphics[scale=0.7]{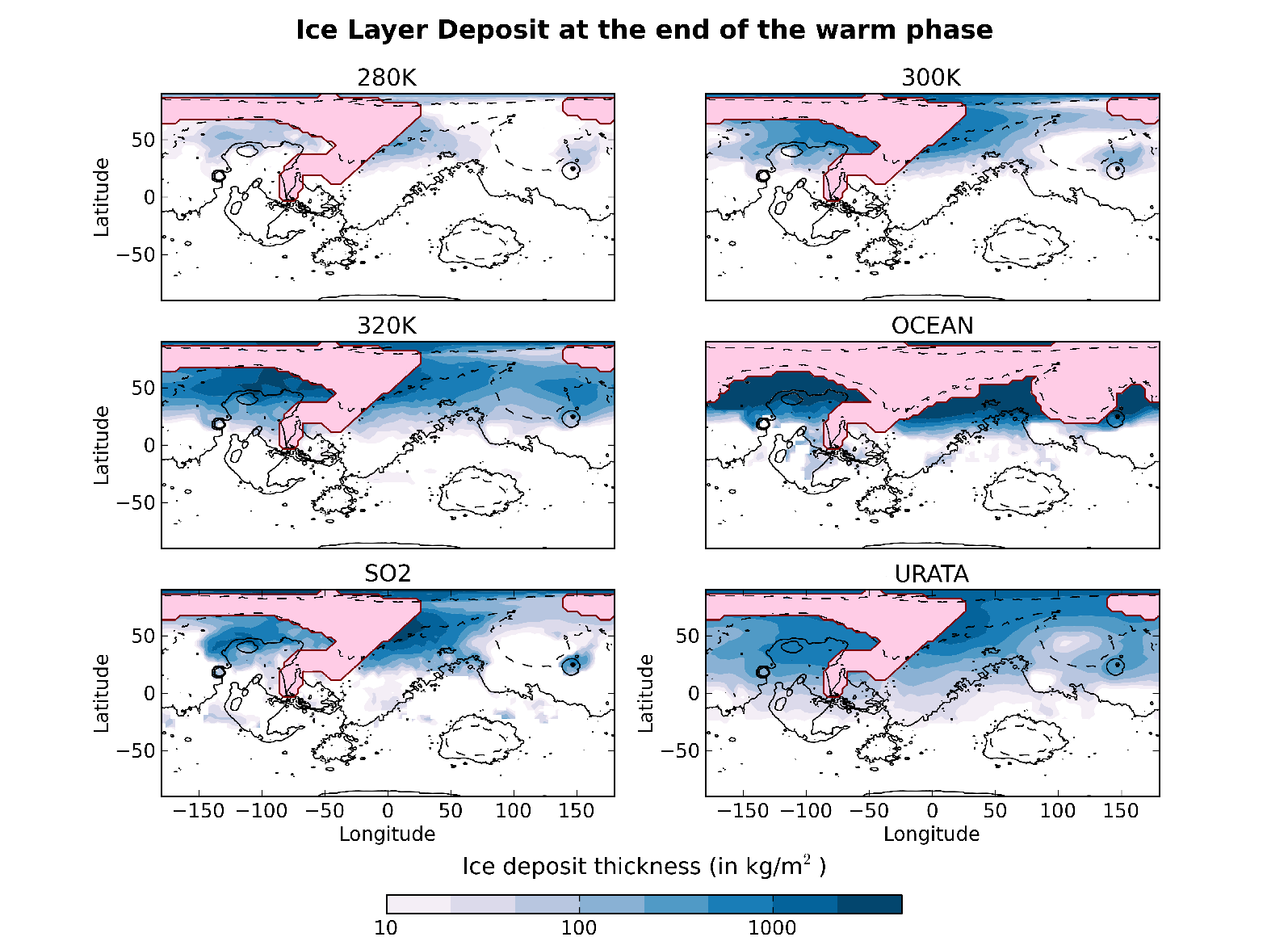}}
\caption{Final Ice layer deposit map (in kg~m$^{-2}$) after 1 martian year of simulations (4 martian years for the ocean case), 
for six different simulations: 1. 280~K outflow, 2. 300~K outflow (reference case), 
3. 320~K outflow, 4. 10$^7$~km$^3$ ocean case, 5. 1~$\%$~SO$_2$ case and 6. l$_0$=$\infty$ (no precipitation case). 
The pink color denotes the regions where the flow passed through on the way to the lake.}
\label{extreme_ice_deposit}
\end{center}
\end{figure*}

\begin{figure*}
\begin{center}
\centerline{\includegraphics[scale=0.7]{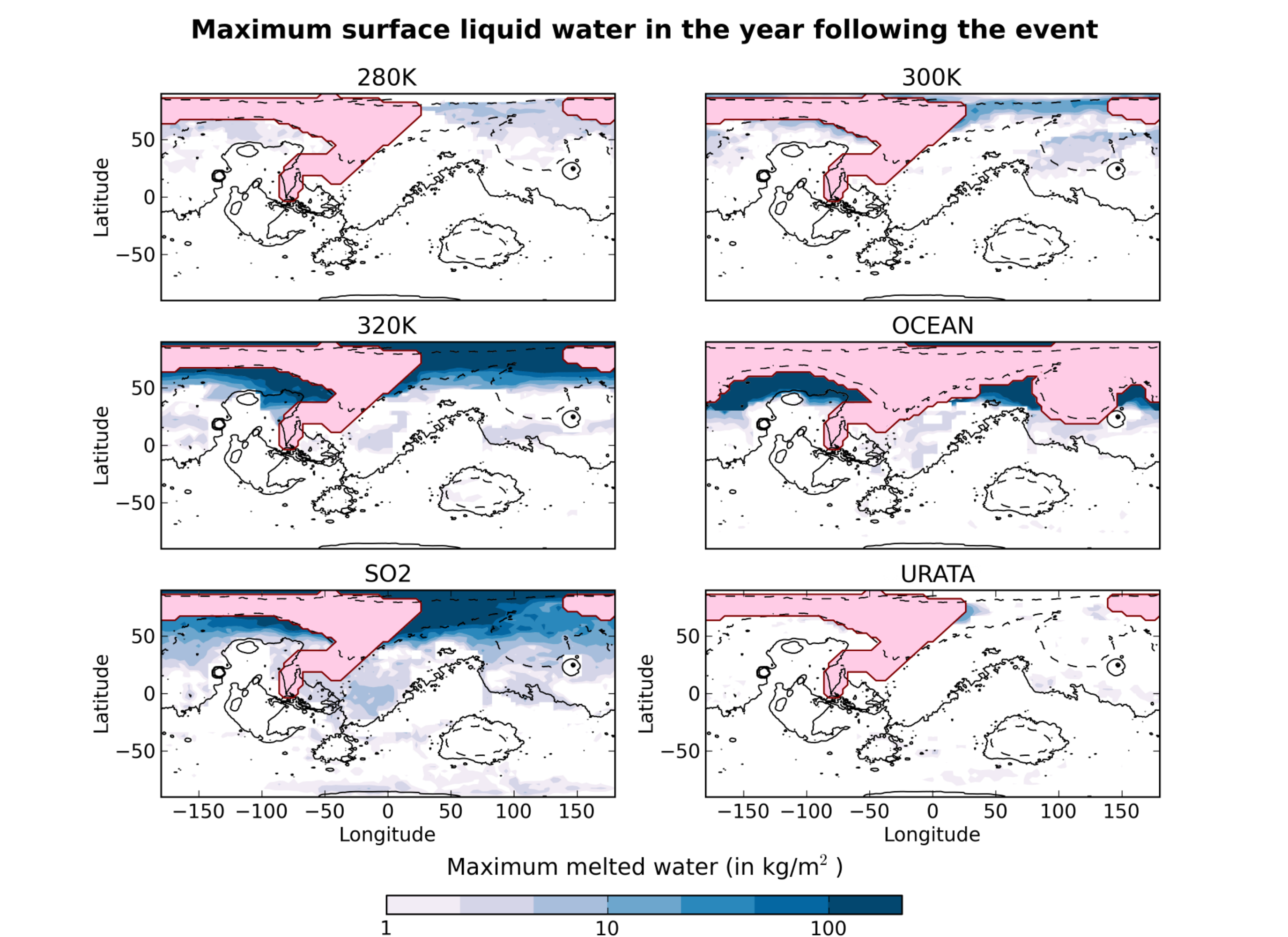}}
\caption{Maximum surface melted liquid water after 1 year of simulation (4 martian years for the ocean case) and for 
6 different simulations: 1. 280~K outflow, 2. 300~K outflow (reference case), 
3. 320~K outflow, 4. 10$^7$~km$^3$ ocean case, 5. 1~$\%$~SO$_2$ case and 6. l$_0$=$\infty$ (no precipitation case).
The pink color denotes the regions where the flow passed through on the way to the lake.}
\label{extreme_max_melted}
\end{center}
\end{figure*}

The amount of water ice transported (Figure~\ref{extreme_ice_deposit}) and melted (Figure~\ref{extreme_max_melted}) 
after outflow channel events with 280~K/300~K/320~K water shows that 
in all cases, the mechanism of advection/cooling to space is very efficient, and as a result, the duration of the warm phase is approximately the same 
($\sim$~500 days) between the reference and the 320~K simulations. 

We note that, at the end of the warm phase, because the amount of ice transported (and the area of the deposit with it) 
increases with the initial temperature of the flow, the average surface albedo raises and the mean temperatures decrease: 
Warmer flows lead to colder states.

\subsubsection{Magnitude of the event: from small outflows to oceans.}

\begin{figure*}
\begin{center}
\centerline{\includegraphics[scale=0.7]{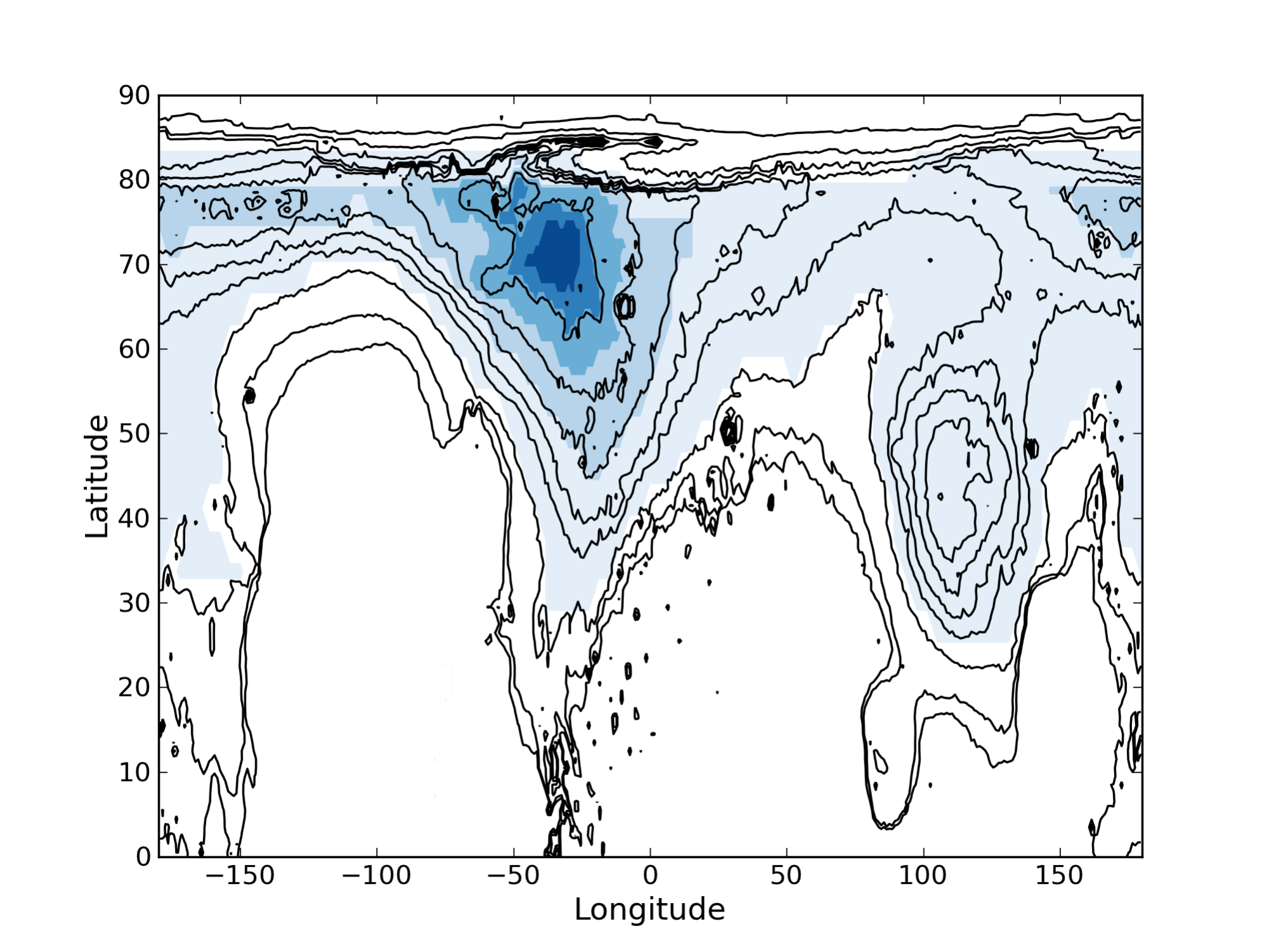}}
\caption{Stable water lake size obtained in our simulations depending on the volume of water released. 
Contour fills are for volumes of 10$^3$,10$^4$,10$^5$,10$^6$ (reference case) and 10$^7$ km$^3$ (ocean case). 
These correspond respectively to wetted areas of 0.15,0.41,1.19,4.15 and 20.4 millions of km$^2$.}
\label{lake}
\end{center}
\end{figure*}

Recent work \citep{Andr:07,Harr:08jgr} has suggested that outflow channels were preferentially carved by multiple events 
of reduced sizes ($\sim$~10$^3$~km$^3$) rather than by large ($>$~10$^5$~km$^3$) single outflows.
We performed simulations for different volumes of water at 300~K and released in Echus Chasma at a rate of 1~km$^3$~s$^{-1}$, 
from 10$^3$~km$^3$ (consistent with the most recent estimations of 
outflow volumes) to 10$^7$~km$^3$ (ocean case). Figure~\ref{lake} shows the final position of the lake as a function of 
the initial volume of water. The 10$^6$~km$^3$ case is the reference simulation.

Our results show that the large outflows, during the warm phase, transport much more water than the small ones (cumulative). 
Small outflows (typically $\sim$~10$^3$-10$^4$~km$^3$) have a small wetted area (typically 0.15-0.41~$\times$~10$^6$~km$^3$) and 
a small initial heat reservoir, so that they cannot warm the atmospheric column above the flow/lake sufficiently to 
transport water vapor into the neighbouring regions. Small outflow events inject more or less the same amount of water vapor (in proportion) 
than large ones, but they are not able to transport it far from the flow/lake. 
For example, 2~$\times$~10$^2$ events of 5~$\times$~10$^3$~km$^3$ transport 2 orders of magnitude less ice outside the flow/lake than 
a large 10$^6$~km$^3$ one (reference simulation). Moreover, large outflows are able to generate precipitation up to $\sim$ 5000~km from 
the edge of the flow/lake whereas small ones cannot produce any precipitation at a distance 
greater than $\sim$~400~km (typically the size of 2 GCM grids). 

We did not explore in detail the effect of the discharge rate, which has a net impact on the size and duration of the wetted area 
(and thus on the evaporation and the albedo), but also on the intensity of the event. 
Nonetheless, the climatic response to lower discharge rate events ($<$~10$^9$~m$^3$~s$^{-1}$) was found to be lower, 
because in such cases the temperatures and the amount of water vapor struggle to build up above the flow/lake.

Because large outflows seem to be much better candidates for generating precipitation globally, 
we examined the extreme case of a catastrophic outflow event 
of 10$^7$~km~$^3$ released simultaneously by all of the circum-Chryse outflow channels (Kasei, Ares, Tiu, Simu Vallis, etc.). 
This possibility, sometimes called the MEGAOUTFLO (Mars Episodic Glacial Atmospheric Oceanic Upwelling by Thermotectonic Flood Outburst) 
hypothesis \citep{Bake:99}, speculates that such events could warm Mars during periods of 10$^4$-10$^5$~years through a transient greenhouse effect 
provoked in part by the injection of large amounts of water vapor. 

\begin{figure}
\begin{center}
\centerline{\includegraphics[scale=0.45]{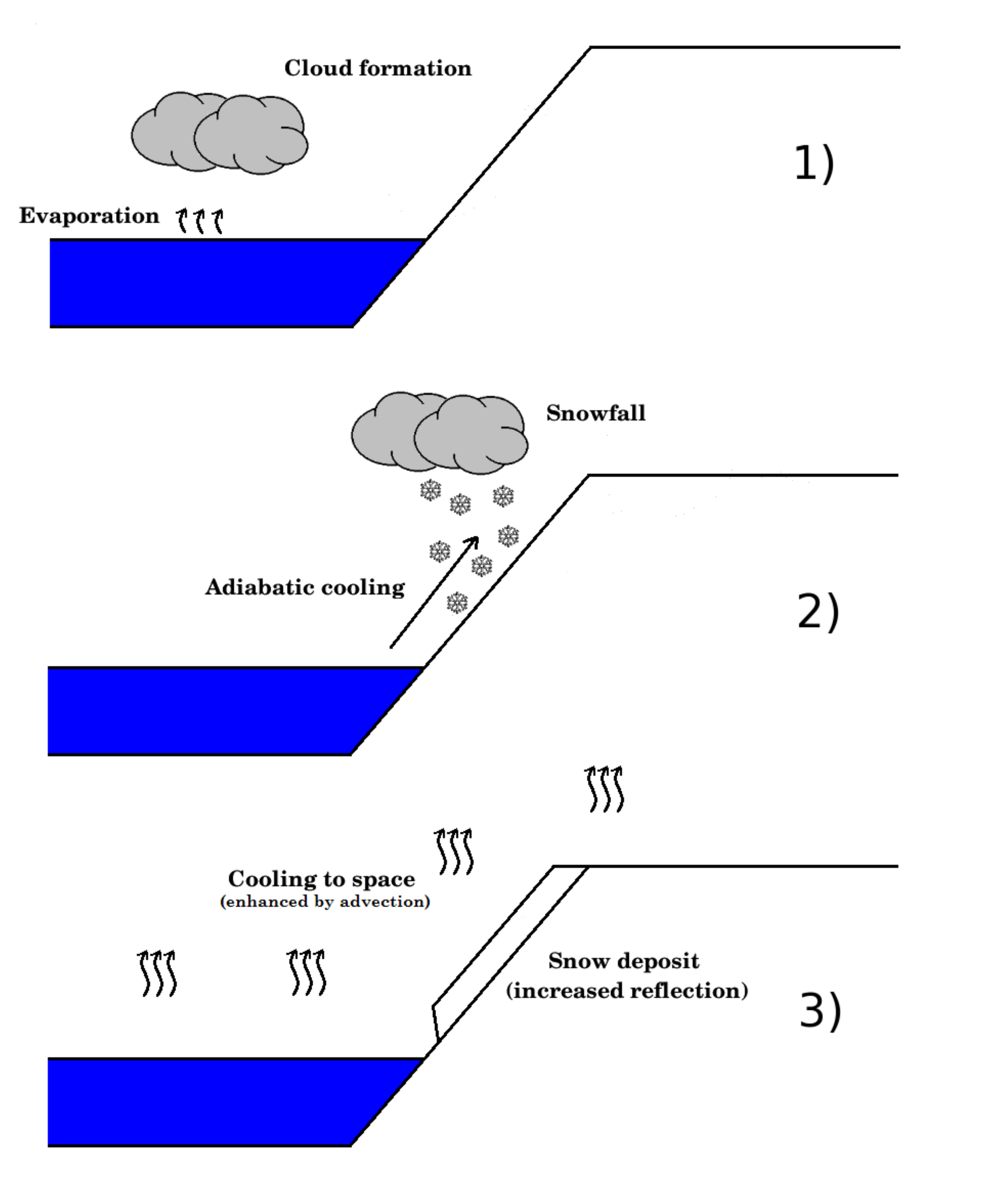}}
\caption{Why is a martian Northern ocean short-lived and unable to induce rain?
The water vapor and the clouds that build up above the ocean (1) are advected southward (2). Because of adiabatic cooling, 
snowfall occurs from the edges of the ocean (starting from 70~$^{\circ}$N, see Figure~\ref{lake}) to the highest parts of the planet 
(the 'Icy Highlands'). The advection of heat (increasing the surface of thermal infrared emission) 
and the increase of reflection (because of the snow deposit) 
both contribute to an intense cooling of the ocean and to reduced precipitation.}
\label{cooling_ocean}
\end{center}
\end{figure}

Our experiments show that such events cannot sustain long-term greenhouse effects, 
whatever the size and the temperatures considered for the northern lake/see/ocean.
After 3.5 martian years, for the outflow event described above, the surface of the lake/see/ocean becomes totally frozen. 
The thermal infrared emission to space (enhanced by the heat horizontal advection 
and by the water vapor advection that release latent heat because of adiabatic cooling; 
see Figure~\ref{cooling_ocean} for the detailed mechanism) acts very efficiently to cool the planet. 
The ice deposited on the Northern Plains slopes (Figure~\ref{extreme_ice_deposit}) also 
enhances the cooling through a depletion of surface solar absorption.
As a result, in such a scenario, rainfall/snowmelt still only occurs in the lowest northern lowlands (see Figure~\ref{extreme_max_melted}) 
of the planet (far from the region of interests).

In summary, the most intense outflow channel events possible are not able to sustain a global greenhouse warming. 
Such events only manage to warm up the atmosphere regionally, in the Northern Plains, and only for a few years at best. 
Consequently, rainfall (and snowmelt) occur only in the neighbourhood regions of the final stable lake. After complete 
surface freezing of the lake, the climate becomes much colder than initially (due to the increase of the surface albedo), 
making the snowmelt even more difficult.

We note that we did not take into account the modification of the topography by the presence of a lake/see/ocean, 
which might be a concern for very high volumes of water ($\geq$~10$^7$~km$^3$).
It could reduce significantly the role of adiabatic cooling and thus favor the transport/deposit of water further south.

\section{Discussion}

\subsection{Role of the atmospheric composition.}

In this analysis, we made the assumption that the Late Hesperian martian atmosphere was made of 100$\%$ CO$_2$ (and 
some water vapor). Outflow channel events under a CO$_2$ dominated atmosphere 
seem not to be able to provoke long-term warming or precipitation at the global scale.

Ouflow channel formation events are very likely related to intense volcanic episodes during martian history \citep{Bake:91,Head:02}. During these periods, 
it is believed that volcanic gases like SO$_2$ may have been massively released [see section~1. of \cite{Kerb:15} for more details].

We performed a simulation of an outflow channel event under the same conditions as in section~\ref{reference_results}, 
but this time with 1~$\%$ of SO$_2$. Figures~\ref{extreme_ice_deposit} and \ref{extreme_max_melted} show 
the corresponding amount of water ice transported/melted after the event.
Small amounts of SO$_2$ (2~mbar here) are sufficient to raise the global atmospheric 
temperatures by several tens of Kelvins and thus to favor the transport 
of water vapor/water ice globally and create precipitation far from the Northern Plains stable lake.

However, using the same GCM, \cite{Kerb:15} (and earlier, \citet{Tian:10}) have shown that massive volcanic SO$_2$ outgassing cannot lead to a global 
and substantial warming, because sulfur aerosols that would form at the same time have a very strong cooling effect, 
even in small amounts.

We also believe that, under more realistic parameterizations that would take into account sulfur aerosols (e.g. \citet{Hale:14}), 
the outflow channel climatic impact would be also very limited. 

\subsection{The role of clouds and precipitation.}
\label{urata_test}

The radiative effect of clouds is one of the main sources of uncertainty in GCMs and thus also on the consistency of our results.

In particular, it has been suggested \citep{Urat:13h2o} that high altitude ('cirrus-like') 
water ice clouds may trigger warm climates on Mars even under a faint young sun.
This scenario requires four assumptions: 
1) Water ice particles that have sizes $>$ 10 microns; 
2) that the rate of precipitation is very low (in order to extend the lifetime of the clouds); 
3) When present, clouds need to completely cover a grid cell (no partial cloud cover); 
4) Lastly, it also requires an initial 'warm' state, for example an outflow channel event.

To explore in a basic manner the role of clouds and precipitation on the climatic impact of outflow channels, we performed a simulation of the reference 
outflow channel event in which we eliminated the precipitation resulting from coalescence (l$_0$=$\infty$). 
For this case, the vertical motion of the ice particles 
is governed only by gravitational sedimentation. Figure~\ref{cloud_coverage} shows that the total cloud cover is near 100$\%$ over all 
the planet during the first year following the event, because of the intense evaporation coupled with the increased lifetime of clouds.

We found that neglecting coalescence and the subsequent precipitation led 
to ice deposits that extend much more areally than in the reference case 
(Figure~\ref{extreme_ice_deposit}), because the lifetime of ice particles increases substantially. 
In such a situation, the global cloud cover (during the year following the event) 
has a net positive radiative impact on the global energy balance of +~12~W~m$^{-2}$ 
(+~21.3~W~m$^{-2}$ of IR warming; -~9.2~W~m$^{-2}$ of solar absorption). 
This is $\sim$+~11~W~m$^{-2}$ higher than in the reference simulation.

However, because the ice field produced by the event extends to a much larger area, the global albedo increases and contributes 
approximately 6~W~m$^{-2}$ of cooling. 
Moreover, because of advection processes, this also increases the horizontal extent of the heat perturbation and thereby the global 
infrared emission to space. Under clear sky conditions, this would lead to an extra cooling of $\sim$~5~W~m$^{-2}$ compared to the reference simulation.

As a consequence, the total rate of cooling is more or less the same ($\sim$~15~W~m$^{-2}$) as that in the reference simulation (l$_0$=0.001).
The duration of the warm phase is also more or less the same than in the reference simulation ($\sim$~500~days).

We also note that the seasonal melting of the deposited ice (see Figure~\ref{extreme_max_melted}) would be very limited 
in such scenarios, because of the increased solar reflection by the clouds. 
In addition, because the ice field produced by the event extends over a large region (Figure~\ref{extreme_ice_deposit}), 
the planet becomes much colder one year after the event than initially.

Nonetheless, we highly encourage further studies 
to explore in more detail the possibility of warming early Mars through water ice clouds (as recently done by \cite{Rami:17}). 

\subsection{Conclusions}

In this analysis, we explored the climatic impact of a wide range of outflow channel events under many possible conditions.

We find that even considering outflow events with intensity (in volumes and temperatures of water released) 
that exceed by far the most recent estimates, the short term climatic response is still very limited. 
The duration of the 'warm' phase that follows the outflow events is completely 
controlled by the total depth and temperature of the lake that is formed and 
is, in practice, no more than few years for the most extreme cases (10$^7$~km$^3$ of water warmed at 300~K, e.g. ocean case). 
In other words, outflow events fail to trigger greenhouse-sustained warm episodes.
Moreover, the precipitation (almost exclusively snowfall) produced by the events during their warm phase 
is limited and confined to the Northern Plains, in the area neighbouring the water outflow.
These results are robust over a wide range of atmospheric pressures and external conditions (e.g. obliquity and season). 

We also find that the intensity of outflow channel event effects can be significantly influenced by 
the atmospheric pressure which is not well constrained for the Hesperian era.
Thin atmospheres (P~$<$~80~mbar), because of their low volumetric heat capacity, can be warmed efficiently.
This can trigger the formation of a convective plume, 
a very efficient mechanism to transport water vapor and ice to the global scale. 
Thick atmospheres (P~$>$~0.5~bar) have difficulty in producing precipitation far from the outflow water locations 
but they are more suited to generate snowmelt.
Nonetheless, outflow channel formation events are unable, 
whatever the atmospheric pressure, to produce rainfall or significant snowmelt at latitudes below 40$^{\circ}$N. 

During the 'cold phase' that follows the solidification to ice of the outflow water, 
the body of water ice emplaced in the Northern Plains has a major contribution to the water cycle. 
The ice is sublimated seasonally and transported progressively 
southward toward the 'Icy Highlands' regions by the processus of adiabatic cooling.
We find that under favorable conditions (obliquity $\sim$~45$^\circ$, atmospheric pressure $\geqslant$ 80~mbar), 
ice deposits can be stabilized in the West Echus Chasma Plateau area.
For an initial 10$^6$~km$^3$ body of water (0.2~bar atmospheric pressure, 45$^\circ$ obliquity), 
they can be present during 10$^5$~martian years.
However, seasonal melting related to solar forcing seems difficult because 
1) the West Echus Chasma Plateau is not ideally located, and 2) the presence of (high albedo) snow at the surface has a significant cooling effect.
The global temperatures after outflow events can thus easily be lowered by few Kelvins making the solar melting possibility even more difficult.
Therefore, in this scenario, localized warming such as geothermal activity or meteoritic impacts 
would be required to explain the formation of valley networks dated 
to the Late Hesperian era and yet observed at this specific location.

\bibliography{biblio_turbet}
\bibliographystyle{apalike}

\end{document}